%% file: paper.tex
\PassOptionsToPackage{pdfpagelabels=false, citecolor=true}{hyperref}
\documentclass[useAMS,usenatbib, usedcolumn]{mnras}
\pdfoutput=1 

\input{Extras/packages}

\input{Extras/commands}

\hypersetup{
    pdfauthor=Victor F. Calderon,
    pdftitle=Prediction of galaxy halo masses in SDSS DR7 via a machine learning approach,
    pdfkeywords={cosmology: observations --- cosmology: large-scale structure 
            of Universe --- galaxies: groups --- galaxies:clusters
            --- methods: statistical, machine learning},
    colorlinks=true,
    citecolor=blue,
    linkcolor=magenta,
    urlcolor=cyan
}

\graphicspath{../Figures/}

\title{ Prediction of Galaxy Halo Masses in SDSS DR7 via a Machine Learning Approach }

\author[Calderon et al. 2018]{
    Victor F. Calderon$^{1}$\thanks{Email: \href{mailto:victor.calderon@vanderbilt.edu}{ victor.calderon@vanderbilt.edu } },
    Andreas A. Berlind$^{1}$\vspace*{0.2em}\\
    $^{1}$ Department of Physics and Astronomy, Vanderbilt University, Nashville, TN 37235, USA
}

\date{Accepted XXX. Received YYY; in original form ZZZ}

\pubyear{ 2018 }

\begin{document}

\setstcolor{red}
\label{firstpage}
\pagerange{\pageref{firstpage}--\pageref{lastpage}}
\maketitle


\input{01_abstract.tex}


\input{02_introduction.tex}


\input{03_data_methods.tex}


\input{04_training_testing.tex}


\input{05_universality_in_model.tex}


\input{06_SDSS.tex}


\input{07_summary_discussion.tex}


\input{08_acknowledgements.tex}


\bibliographystyle{mnras}
\bibliography{Mendeley.bib}


\bsp    
\label{lastpage}
\end{document}

%% file: Extras/packages.tex
\usepackage{ae}
\usepackage{aecompl}
\usepackage{afterpage}
\usepackage{amsmath}
\usepackage{amssymb}
\usepackage{amstext}
\usepackage{appendix}
\usepackage{array}
\usepackage[english]{babel}
\usepackage{blindtext}
\usepackage{booktabs}
\usepackage[markup=underlined, final]{changes}
\usepackage{color}
\usepackage{enumitem}
\usepackage{epstopdf}
\usepackage{float}
\usepackage[T1]{fontenc}
\usepackage{footnote}
\usepackage{footmisc}
\usepackage{graphicx}
\usepackage[figure, figure*]{hypcap}
\usepackage[utf8]{inputenc}
\usepackage{mathptmx}
\usepackage{natbib}
\usepackage{placeins}
\usepackage{relsize}
\usepackage{soul}
\usepackage{tabularx}
\usepackage[most]{tcolorbox}
\usepackage{todonotes}
\usepackage{txfonts}
\usepackage[normalem]{ulem}
\usepackage{units}
\usepackage{url}
\usepackage{xcolor}
\usepackage{xspace}

%% file: Extras/commands.tex
\newcommand{\repourl}{\url{https://github.com/vcalderon2009/SDSS_Groups_ML}}
\newcommand{\catsurl}{\url{http://lss.phy.vanderbilt.edu/groups/ML_Catalogues/}}
\newcommand{\MPAurl}{\url{https://wwwmpa.mpa-garching.mpg.de/SDSS/DR7/}}
\newcommand{\LDurl}{\url{http://lss.phy.vanderbilt.edu/lasdamas/}}
\newcommand{\SKurl}{\url{http://scikit-learn.org/}}
\newcommand{\XGurl}{\url{https://xgboost.readthedocs.io/}}

\newcommand{\refsec}[1]{$\S$\ref{#1}}

\def\calP{{\cal P}}

\def\by{{\bf y}}

\def\bF{{\bf F}}

\def\bX{{\bf X}}

\newcommand{\MPA}{MPA-JHU }
\newcommand{\LD}{\texttt{LasDamas}}

\newcommand{\XGBoost}{\texttt{XGBoost}}
\newcommand{\XGBoosts}{\texttt{XGBoost} }
\newcommand{\RF}{\texttt{RF}}
\newcommand{\RFs}{\texttt{RF} }
\newcommand{\NN}{\texttt{NN}}
\newcommand{\NNs}{\texttt{NN} }
\newcommand{\sklearn}{\texttt{scikit-learn}}
\newcommand{\MLP}{\texttt{MLP}}
\newcommand{\HAM}{\texttt{HAM}}
\newcommand{\HAMs}{\texttt{HAM} }
\newcommand{\DYN}{\texttt{DYN}}
\newcommand{\DYNs}{\texttt{DYN} }

\newcommand{\figwidth}{0.47}

\def\beq{\begin{equation}}
\def\eeq{\end{equation}}
\def\barray{\begin{eqnarray}}
\def\earray{\end{eqnarray}}
\def\beqarray{\begin{eqnarray}}
\def\eeqarray{\end{eqnarray}}

\def\gtsima{$\; \buildrel > \over \sim \;$}
\def\ltsima{$\; \buildrel < \over \sim \;$}
\def\prosima{$\; \buildrel \propto \over \sim \;$}
\def\gsim{\lower.7ex\hbox{\gtsima}}
\def\lsim{\lower.7ex\hbox{\ltsima}}
\def\simgt{\lower.7ex\hbox{\gtsima}}
\def\simlt{\lower.7ex\hbox{\ltsima}}
\def\simpr{\lower.7ex\hbox{\prosima}}
\def\la{\lsim}
\def\ga{\gsim}

\newcommand{\sig}[1]{$#1\sigma$}
\newcommand{\fracdm}{\Delta f}
\newcommand{\fracd}{$\fracdm$}
\newcommand{\fracds}{$\fracdm$ }

\newcommand{\dvm}{\sigma_{v,b}}

\newcommand{\dvs}{$\dvm$ }
\newcommand{\fvbm}{f_\mathrm{vb}}
\newcommand{\fvb}{$\fvbm$}
\newcommand{\fvbs}{$\fvbm$ }
\newcommand{\Mhamm}{M_{\textrm{HAM}}}

\newcommand{\Mhams}{$\Mhamm$ }
\newcommand{\Mdynm}{M_{\textrm{dyn}}}
\newcommand{\Mdyn}{$\Mdynm$}
\newcommand{\Mdyns}{$\Mdynm$ }

\setstcolor{red}
\definecolor{orange}{rgb}{1,0.5,0}
\setremarkmarkup{\todo[color=Changes@Color#1!20,size=\scriptsize]{#1: #2}}
\newcommand{\note}[2][]{}

\newcommand{\msunh}{\>h^{-1}\rm M_\odot}

\newcommand{\mgroup}{\ensuremath{M_\textrm{group}}\xspace}

\newcommand{\msun}{h^{-1}M_\odot}

\newcommand{\hmpc}{\mbox{$h^{-1}\textrm{Mpc}$} }
\newcommand{\kmsMpc}{\mbox{$\textrm{km}\ \textrm{s}^{-1}\ \textrm{Mpc}^{-1}$}}
\newcommand{\hmpcthreeinv}{\mbox{$h^{3} \textrm{Mpc}^{-3}$}}

\newcommand{\gr}{\ensuremath{(g-r)}\xspace}

\newcommand{\grbs}{$\mathbf{(g-r)}$ }
\newcommand{\ssfr}{{sSFR}}
\newcommand{\ssfrs}{{sSFR }}
\newcommand{\ssfrb}{\textbf{{\ssfr}\xspace}}
\newcommand{\sfr}{{SFR}\xspace}

\newcommand{\sersic}{{S\'ersic}\xspace}

\newcommand{\rband}{\textit{r}-band}
\newcommand{\rbands}{\textit{r}-band }

\newcommand{\gbands}{\textit{g}-band }

\newcommand{\MD}[1]{{\texttt Mr#1-SDSS}\xspace}

\newcommand{\kcor}{\textit{k}-corrected }
\newcommand{\GRrms}{$\mathbf{R_{\perp,rms}}$}
\newcommand{\Gdv}{$\mathbf{\sigma_{v}}$}

\newcommand{\Mrtotb}{$\mathbf{M_{r,\textrm{tot}}}$}
\newcommand{\ngal}{$n_\mathrm{gal}$}
\newcommand{\wprp}{$w_{p}\left(r_p\right)$}
\newcommand{\mpredm}{M_{\textrm{pred}}}
\newcommand{\mtruem}{M_{\textrm{true}}}
\newcommand{\mpred}{$\mpredm$}
\newcommand{\mpreds}{$\mpredm$ }
\newcommand{\mtrue}{$\mtruem$}

\newcommand{\deltalogmm}{(\Delta\mathrm{log}M)_{68}}
\newcommand{\deltalogm}{$\deltalogmm$}
\newcommand{\deltalogms}{$\deltalogmm$ }
\newcommand{\sigloglm}{\sigma_{\mathrm{log}L}}
\newcommand{\siglogl}{$\sigloglm$}


%% file: 01_abstract.tex
\begin{abstract}

We present a machine learning (ML) approach for the prediction of
galaxies' dark matter halo masses that achieves an improved performance
over conventional methods. We train three ML algorithms (\XGBoost,
Random Forests, and neural network) to predict halo masses using a
set of synthetic galaxy catalogues that are built by populating dark
matter haloes in N-body simulations with galaxies, and that match both
the clustering and the joint-distributions of properties of galaxies in
the Sloan Digital Sky Survey (SDSS). We explore the correlation of
different galaxy- and group-related properties with halo mass, and extract
the set of nine features that contribute the most to the prediction of
halo mass. We find that mass predictions from the ML algorithms are
more accurate than those from halo abundance matching (\HAM) or
dynamical mass (\DYN) estimates. Since the danger of this approach is that
our training data might not accurately represent the real Universe, we
explore the effect of testing the model on synthetic catalogues
built with different assumptions than the ones used in the training phase.
We test a variety of models with different ways of populating dark
matter haloes, such as adding velocity bias for satellite galaxies.
We determine that, though training and testing on different data can lead
to systematic errors in predicted masses, the ML approach still
yields substantially better masses than either \HAMs or \DYN. Finally,
we apply the trained model to a galaxy and group catalogue from the SDSS DR7
and present the resulting halo masses.
\end{abstract}

%% file: 02_introduction.tex

\section{Introduction}
\label{sec:intro}

The practice of grouping galaxies observed in a galaxy catalogue
into galaxy groups and clusters has been utilised extensively in
astrophysics and cosmology, since the pioneering work of George
Abell and Fritz Zwicky \citep{Abell1958,Zwicky1968}, who constructed
cluster catalogues from the Palomar Observatory Sky Survey (POSS) 
using local galaxy surface number densities. Galaxy clusters
represent the largest primordial density perturbations to have formed 
by now, and typically contain tens to hundreds of galaxies embedded 
within a common dark matter halo\footnote{Throughout this paper, we use 
the term "halo" to refer to a gravitationally bound structure with
overdensity $\rho/\bar{\rho} \sim 200$, so an occupied halo may host a
single luminous galaxy, a group of galaxies, or a cluster.}, thus tracing 
the high mass tail of the halo mass function. As a result, clusters 
constitute one of the most powerful cosmological probes and measurements 
of their abundance can be used to constrain cosmological parameters 
\citep[e.g.,][]{Voit2005,Allen2011,Kravtsov2012,Weinberg2013,Mantz2014}.
Additionally, our current understanding of galaxy formation and evolution
revolves around the idea that all galaxies are formed and live within 
dark matter haloes. Therefore, galaxy groups and clusters, if identified 
correctly, can be used to study the \textit{galaxy-halo} connection 
and thus how galaxies form and evolve within dark matter haloes.
Whether we wish to use galaxy groups as probes of cosmology or galaxy
formation, determining their masses accurately and robustly has proven to 
be a difficult task.

Galaxy groups and clusters can be identified in various ways. Originally, 
clusters were first detected as overdensities of galaxies in broad-band 
images in the visible spectrum \citep[e.g.][]{Abell1958,Zwicky1968}.
Since then, clusters have mainly been identified as overdensities
of red galaxies in visible and IR bands 
\citep[e.g.][]{Gladders2005,Hao2010,Ascaso2012}, as extended X-ray sources 
\citep[e.g.][]{Rosati2002,Vikhlinin2009}, or by their signature in the 
cosmic microwave background
\citep[e.g.][]{Marriage2011,Staniszewski2009,PlanckCollaboration2015}.
Since the early 1980's and with the onset of redshift surveys, groups of 
galaxies have also been selected based on the closeness of galaxies in 
redshift space using three-dimensional algorithms. Many of these analyses 
have adopted the widely-used Friends-of-Friends percolation algorithm 
\citep{Geller1983a} to place galaxies into groups and thus compile group 
catalogues. This algorithm links galaxies in pairs based on their 
separation along the line-of-sight or on the sky and places all linked
galaxies into a single group. Numerous group galaxy catalogues have been 
constructed in this way for different redshift surveys, including the 
Center for Astrophysics Redshift Survey \citep[CfA;][]{Geller1983a}, the 
Las Campanas Survey \citep{Tucker2000}, the Two Degree Field Galaxy 
Redshift Survey 
\citep[2dFGRS; ][]{Merchan2002,Eke2004,Yang2005,Tago2006,Einasto2007},
the high-redshift DEEP2 survey \citep{Gerke2005}, the Two Micron
All Sky Redshift Survey \citep{Crook2007}, and the Sloan Digital
Sky Survey \citep[e.g.,][]{Goto2005,Berlind2006}.


Once galaxy groups and clusters are identified, mass measurements are
needed to map observable properties to the underlying masses of dark
matter haloes. Traditionally, there are two main methods to assign masses 
to galaxy groups and clusters that are built from galaxy redshift surveys, 
i.e., Halo Abundance Matching \citep[hereafter \HAM; e.g.,][]{Kravtsov2004,Tasitsiomi2004,Vale2004,Conroy2006} and dynamical 
mass estimates \citep[hereafter \DYN; e.g.,][]{Teague1990,Colless1996,Fadda1996,Carlberg1997,Girardi1998,Brodwin2010,Rines2010,Sifon2013,Ruel2014}.
The \HAMs method assumes a monotonic relation between a theoretical 
mass-like quantity related to dark matter haloes and another observable 
quantity related to galaxies. This approach is simple yet powerful, wherein 
matching cumulative number densities of galaxies and haloes yields an 
implicit relationship between the theoretical quantity and the 
observational quantity \citep{Hearin2013}. \HAMs is typically used to 
connect galaxies to both host haloes and subhaloes, but in this context we 
refer to a variant of the method that connects galaxy groups to host
haloes alone. For example, \citet{Yang2007} 
applied a halo-based group-finder \citep{Yang2005} to the 2dFGRS and 
assigned halo masses to galaxy groups based on characteristic luminosity 
and characteristic stellar mass. \citet{Lim2017} extended this approach and 
applied a modified version of the same algorithm to multiple large redshift 
surveys. \citet{Calderon2018} applied the \citet{Berlind2006} algorithm
to the SDSS and used \HAMs to estimate halo masses, based on the integrated
luminosity of the groups. \citet{Moffett2015} did the same for the REsolved 
Spectroscopy of a Local VolumE \citep[RESOLVE;][]{Eckert2015} and the 
Environmental COntext catalog \citep[ECO;][]{Moffett2015}.
On the other hand, \DYNs estimates of clusters use the line-of-sight 
velocity dispersion of galaxies within clusters, together with measurements
of their size, as dynamical tracers of the underlying gravitational 
potential. These estimates make use of variants of the virial theorem to 
estimate group masses.

Each of these approaches are not perfect, and may include possible biases
or systematic errors in their mass estimates that may influence the final
results. \citet{Old2014} performed an extensive comparison between various 
galaxy-based cluster mass estimation techniques that use position, 
velocities, and colours of galaxies to quantify the scatter, systematic 
biases and completeness of cluster masses derived from a diverse set of 25 
galaxy-based methods. They found that abundance-matching and richness-based 
methods provide the best results, with some estimates being under- and 
overestimated by a factor greater than ten. \citet{Wojtak2018} studied 
these results further and found that contamination in cluster membership can
affect the mass estimates greatly, with all methods either overestimating 
or underestimating the final cluster masses when applied to contaminated 
or incomplete galaxy samples, respectively. Additionally, 
\citet{Armitage2018} used the \texttt{C-EAGLE} galaxy clusters sample 
\citep{Barnes2017} to quantify the bias and scatter of three mass 
estimators, and found no significant bias, but a large scatter when
comparing estimated to true masses. For the case of \HAMs, 
\citet{Campbell2015d} compared three different FoF-based group-finding 
algorithms by applying them to a realistic mock galaxy catalogue where the
halo masses are known. They found that estimating group masses using \HAMs 
is limited by the intrinsic scatter in the relation between the observed 
quantity and the halo mass. They also show that errors in the
group-finding process can cause catastrophic errors in estimated halo mass.

These previous works have demonstrated that galaxy groups and clusters 
identified in redshift surveys have mass estimates that are prone to large 
statistical and systematic errors, mostly due to failures of the group
finding algorithms. These methods for estimating mass use one or two 
properties of groups, such as total luminosity in the case of \HAM, or
velocity dispersion and radius in the case of \DYN. However, there are
many additional properties of groups that should contain information about
halo mass, such as colours and star formation rates, full density and 
velocity profiles, large scale environments, etc. This suggests the 
opportunity to apply nonparametric algorithms to analyse the abundant data 
at our disposal. There has been a significant increase in recent years in 
the number of studies applying machine learning (ML) techniques to 
astronomy. One of the most important applications of ML in astronomy is the 
classification of various objects, e.g. transient events 
\citep{Mahabal2008} and galaxy morphology \citep{Banerji2010}. Other 
applications include the determination of photometric redshifts of 
galaxies from a set of broadband filters \citep{Ball2007,Gerdes2010}, the 
assignment of dark matter haloes to generate synthetic catalogues from 
N-body simulations \citep{Xu2013,Kamdar2016,Kamdar2016a} and the study of 
the structure of the Milky Way \citep{Riccio2016}. Relevant to this work,
ML has also been used to improve galaxy cluster dynamical mass measurements 
by employing the entire line-of-sight velocity PDF information of galaxy 
clusters \citep{Ntampaka2015,Ntampaka2016}. More recently, ML algorithms 
have been used to measure cluster masses using a combination of dynamical 
and X-ray data \citep{Armitage2018b}, and more complex algorithms have 
been employed to estimate the masses of galaxy clusters using synthetic 
X-ray images from cosmological simulations \citep{Ntampaka2018}.
However, these studies were restricted to the massive cluster regime.

In this paper, we explore the possibility of employing ML techniques to 
estimate the halo masses of galaxies in a wide range of mass. We adopt
observed properties of both the galaxies and their groups to act as
features and we train the ML algorithms on synthetic data.
This paper is organised as follows. In \refsec{sec:data_meas},
we describe the observational and simulated data used in this
work (\refsec{subsec:SDSS_galaxy_catalogues}), introduce the set of
\textit{features} used in this analysis (\refsec{subsec:galprop_features}),
and present the main set of ML algorithms that we use 
(\refsec{subsec:ml_algorithms}). In \refsec{sec:training_testing},
we provide the main analysis of \textit{feature selection}
(\refsec{subsec:feat_selection}), and present our main results
of mass estimates (\refsec{subsec:fixed_hod_dv}). In \refsec{sec:universality}
we also present a detailed examination of how mass estimates may vary
depending on the choice of HOD parameters (\refsec{subsec:varying_hod}),
velocity bias, \dvs (\refsec{subsec:varying_dv}), or scatter in the 
mass-to-light ratio of central galaxies (\refsec{subsec:varying_scatter}).
In \refsec{sec:applying_sdss}, we
apply our trained algorithms to SDSS, and present the resulting galaxy
catalogue with various estimates of halo mass.
We summarise our results and discuss their implications in
\refsec{sec:summary_discussion}. The Python code and catalogues
used in this project will be made publicly available on
Github\footnote{\repourl} upon publication of this paper.

%% file: 03_data_methods.tex
\section{Data and Methods}
\label{sec:data_meas}

In this section, we present the datasets used throughout this analysis,
and introduce the main ML algorithms and statistical
methods that we use to estimate the halo masses of galaxies. 
In \refsec{subsec:SDSS_galaxy_catalogues}, we briefly describe
the SDSS galaxy sample and synthetic galaxy catalogues that we use,
along with the parameters that are included in these catalogues. In
\refsec{subsec:galprop_features}, we introduce the different
\textit{features} that we use for training our ML predictors,
and provide a guide on how these are calculated. Finally, in 
\refsec{subsec:ml_algorithms} we provide a brief overview of the
different algorithms that we use in this analysis, as well as the
default tuning parameters used by each algorithm.
\subsection{SDSS Galaxy Sample and Mock Galaxy Catalogues}
\label{subsec:SDSS_galaxy_catalogues}

For this analysis, we make use of a modified version
of the galaxy and group galaxy catalogues used in
\citet{Calderon2018}. We will provide a brief description of
the galaxy sample used, and also an overview of the synthetic
galaxy and group galaxy catalogues used in this analysis.

\subsubsection{SDSS Galaxy Sample}
\label{subsubsec:NYU_DR7}

For this analysis, we use data from the Sloan Digital
Sky Survey \citep[hereafter SDSS;][]{York2000b}.
SDSS collected its data with a dedicated
2.5-meter telescope \citep{Gunn2006}, camera
\citep{Gunn1998}, filters \citep{Doi2010}, and spectrograph
\citep{Smee2012}. We construct our galaxy sample from
the {\texttt{large-scale structure}} sample of the
NYU Value-Added Galaxy Catalogue \citep[NYU-VAGC;][]{Blanton2005},
based on the spectroscopic sample in Data Release 7
\citep[SDSS DR7;][]{Abazajian2008}. The main
spectroscopic galaxy sample is approximately complete
down to an apparent \rband\ Petrosian magnitude limit of
$m_{r} = 17.77$. However, we have cut our sample back to
$m_{r} = 17.6$ so it is complete down to that 
magnitude limit across the sky. Galaxy absolute magnitudes
are \textit{k}-corrected \citep{Blanton2003a} to rest-frame
magnitudes at redshift $z=0.1$.

We construct a volume a volume-limited galaxy sample
that contains all galaxies more luminous than $M_{r} = -19$,
and we refer to this sample as \MD{19}. The redshift
limits of the sample are $z_\mathrm{min}=0.02$ and
$z_\mathrm{max}=0.067$ and it contains 90,893 galaxies with
a number density of $n_\mathrm{gal} = 0.01503$ \hmpcthreeinv.
The sample includes the right ascension, declination, redshift,
and \gr colour for each galaxy.

To each galaxy, we assign a star formation rate (\sfr) using
the \MPA Value-Added Catalogue DR7\footnote{\MPAurl}.
This catalogue includes, among many other parameters, stellar
masses based on fits to the photometry using \citet{Kauffmann2003a}
and \citet{Salim2007}, and star formation rates based on
\cite{Brinchmann2004}. We cross-match the galaxies of the
NYU-VAGC to those in the \MPA catalogue using their
MJD, plate ID, and fibre ID. A total of 5.65\% of galaxies
in the sample did not have corresponding values of \sfr
and were removed from the main sample. This leaves a sample
of 85,578 galaxies. For each of these galaxies, we divide its
\sfr by its stellar mass to get specific star formation
rates, \ssfr.

Ultimately, we identify galaxy groups using the 
\citet{Berlind2006} group-finding algorithm. This is a
Friends-of-Friends \citep[FoF;][]{Huchra1982} algorithm
that links galaxies recursively to other galaxies that are
within a cylindrical linking volume. The projected and
line-of-sight linking lengths are $b_{\perp} = 0.14$ and
$b_{\parallel} = 0.75$ in units of the mean inter-galaxy
separation, respectively. This choice of linking lengths was
optimised by \citet{Berlind2006} to identify galaxy systems that
live within the same dark matter halo. In each group, we define
the most luminous galaxy (in the \rband) to be the
'central' galaxy. The rest of the galaxies are defined as
'satellite' galaxies. 

In previous works, we have estimated the total masses of the 
groups via \textit{abundance matching}, using total group luminosity
as a proxy for mass. Specifically, we assume
that the total group \rband\ luminosity $L_{\textrm{group}}$
increases monotonically with halo mass $M_{\textrm{h}}$, and
we assign masses to groups by matching the cumulative space
densities of groups and haloes:
\begin{align}
n_{\textrm{group}} (> L_{\textrm{group}}) &=
n_{\textrm{halo}}(> M_{\textrm{h}}).
\end{align}
To calculate the space densities of haloes, we adopt the
\cite{Warren2006} halo mass function assuming a cosmological
model with $\Omega_{m} = 1 - \Omega_{\Lambda} = 0.25$,
$\Omega_{b} = 0.04$, $h \equiv H_{0}/$ (100 \kmsMpc) = 0.7,
$\sigma_{8} = 0.8$, and $n_{s} = 1.0$. We refer to these
abundance matched masses as \textit{group masses}, \mgroup.
In this paper, we also use a dynamical mass estimate for
each group, as well as other group properties, which are
described in \S~\ref{subsec:galprop_features}.

\subsubsection{Mock Galaxy Catalogues}
\label{subsubsec:Mock_Catls}

In order to make proper predictions of the halo masses
of galaxies, we need a training dataset where the halo 
mass of each galaxy is known. This necessitates that we
use mock, rather than real data. However, the accuracy of
our predictions hinges on the degree to which the mock 
data are truly representative of the observable Universe. 
Therefore, the mock dataset must not only contain the same
observable properties that we will use as features in the 
SDSS data, it should also faithfully reproduce the true 
correlations between these properties and halo mass. At a 
minimum, the training data should be able to accurately 
reproduce the observed clustering of galaxies and the 
joint distributions of "observed" galaxy properties.

For this project, we use a suite of 10 realistic
synthetic galaxy and group galaxy catalogues similar to
\citet{Calderon2018}, with the one exception that we use a 
different definition when identifying dark matter haloes, 
i.e. we use a \textit{spherical-overdensity} (SO) definition 
as opposed to the \textit{Friends-of-Friends} (FoF) halo 
definition used in \citet{Calderon2018}. These synthetic 
catalogues are based on the \textit{Large Suite of Dark Matter 
Simulation} (\LD) project\footnote{\LDurl} \citep{McBride2009}, 
and have the same clustering and same distributions of "observed" 
properties as the SDSS data (luminosity, \gr colour, \ssfr, and 
\sersic index). We use an Halo Occupation Distribution 
\citep[HOD;][]{Berlind2002} model to populate the DM haloes with 
central and satellite galaxies, whose numbers as a function of 
halo mass were chosen to reproduce the number density, \ngal, 
projected 2-point correlation function, \wprp, and group
multiplicity function, $n(N)$, of the \MD{19} sample. Specifically, 
we use the best-fit HOD values of \citet{Sinha2018} for the case 
of the \MD{19} sample, the `LasDamas' cosmology, the `Mvir' halo 
definition, and the `PCA' option.

Once galaxies are placed in haloes, we assign luminosities and 
colours using modified versions of the Conditional Luminosity 
Function \citep[CLF;][]{Yang2003} framework and the \citet{Zu2016} 
halo-quenching model. This approach yields luminosity and colour 
distributions as well as luminosity- and colour-dependent clustering 
that are in agreement with SDSS measurements.
The resulting mock catalogues have been analysed in exactly the same 
way as the SDSS data (i.e. same group-finding algorithm, same method 
of assigning group masses, etc). In their final version, the 
catalogues contain information on various galaxy-related properties 
(e.g., \ssfr, \sersic index, \gr colour, luminosity) and group-related
properties (e.g., group richness, groups' total \rband\ absolute 
magnitudes, velocity dispersion within the groups, etc).

For a more detailed explanation of what went into producing
the set of mock catalogues used in this analysis, we refer
the reader to \S 2.3 of \citet{Calderon2018}.

\subsection{Galaxy properties as \textit{features}}
\label{subsec:galprop_features}

As part of our analysis, we must make a decision on which
\textit{features} to use when training the ML algorithms to 
predict the masses of galaxies' dark matter halos. The set 
of features that we use includes properties of the
galaxy in question as well as properties of the group to 
which the galaxy belongs. All features can be observed and 
measured in the SDSS. Here we provide a list of the features 
that we consider initially with a description of how each 
is computed. Later on we reduce this to a shorter list 
using a feature selection algorithm. 

\paragraph*{Galaxy-related features}
\begin{enumerate}[label=\arabic*, leftmargin=1\parindent, labelsep=0.5\parindent]
\itemsep0.5em
\item \textbf{Distance to group's centre}:
    This feature refers to how far a galaxy is from the 
    centre of its corresponding galaxy group. This variable
    is given in units of of \hmpc, but it is calculated in 
    three-dimensional space so it is dominated by the velocity
    component of the galaxy's position. The centre of the group 
    is computed as the centroid of the group's member galaxy 
    positions.
\item \textbf{Absolute Magnitude}:
    \rbands absolute magnitude of the galaxy, \textit{k}-corrected 
    to $z=0.1$.
\item \textbf{Specific star formation rate of the galaxy}, \ssfrb:
    Logarithmic value of the specific star formation rate of the
    galaxy. As mentioned in \refsec{subsubsec:Mock_Catls} and in
    \citet{Calderon2018}, in our mock catalogues these \ssfr 
    values were assigned using the \citet{Zu2016} 
    \textit{halo-quenching} model, and matched to the distribution of \ssfrs values in SDSS DR7 through abundance matching.
\item \textbf{Group galaxy type}:
    The galaxy type of the galaxy, in terms of its galaxy group.
    We denote a value of "1" if the
    galaxy is a \textit{group central}, and a "0" if the galaxy is
    a \textit{group satellite}. After determining the group 
    membership of each galaxy, we designate the brightest
    galaxy of the group in the \rbands as the \textit{group central},
    while the rest of galaxies are identified as 
    \textit{group satellites}. Hence, a galaxy group is composed of one
    bright group central and a number of group satellites. This criterion
    is motivated by the idea that central galaxies grow in mass and
    brightness by galactic cannibalism \citep{Dubinski1998,Cooray2005},
    while satellite galaxies experience a series of events that strip
    them from their mass and inhibit star formation (e.g. ram-pressure
    stripping and tidal stripping).
\item \textbf{\grbs colour of galaxy}:
    The difference between the absolute magnitudes in the 
    \gbands and \rband, after these have been \kcor to $z=0.1$.
    In our mock catalogues, galaxy colours were assigned
    in a manner similar to that of sSFR.
\end{enumerate}

\paragraph*{Group-related features}
\begin{enumerate}[label=\arabic*, leftmargin=1\parindent, labelsep=0.5\parindent]
\setcounter{enumi}{5}
\itemsep0.5em
\item \textbf{Luminosity of brightest galaxy}:
    \rband\ absolute magnitude value of the brightest galaxy in the group
    that the galaxy in question belongs to.
    This absolute magnitude is the same as that of the group central
    galaxy, according to our designation of \textit{group centrals}
    and \textit{group satellites}.
\item \textbf{Luminosity ratio}:
    Ratio between the \rbands luminosity of the brightest
    and second brightest galaxies in the group.
\item \textbf{Total luminosity}, \Mrtotb:
    The total \rband\ luminosity of the group is the sum of the
    \rbands luminosities of its member galaxies. We compute the total
    group \rbands absolute magnitudes as
    \begin{align}
        M_{r,\textrm{tot}} = -2.5\log_{10} \left ( \sum^{N}_{i=1} 10^{-0.4 M_{0.1_{r,i}}} \right ),
    \end{align}
    where `$N$` corresponds to the number of member galaxies in the
    group, and `$M_{0.1_{r,i}}$` to the \kcor \rband\ absolute magnitude
    of the \textit{i}-th galaxy in the galaxy group. The resulting
    variable is the groups' total \rbands absolute magnitude, 
    $M_{r,\textrm{tot}}$.
\item \textbf{Total specific star formation rate}, $\mathbf{\ssfr_{G}}$:
    Logarithmic value of the total specific star formation rate of
    the group. For each group, the total specific star formation rate
    is calculated as:
    \begin{align}
        \textrm{\ssfr}_{G} &= \frac{\textrm{SFR}_{G}}{M_{*,G}} = \frac{\displaystyle \sum\limits^{N}_{i = 1} \textrm{SFR}_{i}}{ \displaystyle \sum\limits^{N}_{i = 1} M_{*,i} },
    \end{align}
    where `$N$` refers to the number of member galaxies in the
    galaxy group, `$M_{*,i}$` and `$\textrm{SFR}_{i}$` to the stellar mass
    and star formation rate of the \textit{i}-th galaxy in the
    galaxy group.
\item \textbf{Shape}:
    The shape of the group is calculated by first computing the eigenvalues of the group's moment of inertia tensor, and then
    by taking the ratio between the values of the largest and second largest eigenvalues. This ratio is what we designate as the \textit{group shape} feature.
\item \textbf{Richness}:
    Richness is the total number of galaxies in the
    galaxy group. A galaxy group can be composed of a single
    galaxy, or many galaxies.
\item \textbf{Projected rms radius, \GRrms}:
    Projected \textit{rms} radius of the group. It is given by
    \begin{align}
        R_{\perp,rms} = \sqrt{\frac{1}{N} \sum^{N}_{i=1} r_{i}^2},
    \end{align}
    where $r_i$ is the projected distance between each member galaxy
    and the group centroid. This variable is only computed for
    galaxy groups with two or more member galaxies. For groups with
    just one member galaxy, we assign a value of '0' to \GRrms.
\item \textbf{Maximum projected radius}, $\mathbf{r_{\mathrm{\textbf{tot}}}}$:
    The total radius of the galaxy group corresponds to
    the projected distance between the centre of the galaxy group and
    and the most distant member galaxy of the group.
\item \textbf{Median projected radius}, $\mathbf{r_{\mathrm{\textbf{med}}}}$:
    The median radius of the galaxy group is the median distance 
    between the centre of the group and the group's member galaxies.
\item \textbf{Total velocity Dispersion}, \Gdv:
    We compute a group one-dimensional velocity dispersion given by
    \begin{align}\label{eq:group_vel_dispersion}
        \sigma_{v} = \frac{1}{1 + \bar{z}} \sqrt{\frac{1}{N - 1} \sum^{N}_{i = 1} (cz_{i} - c\bar{z})^2},
    \end{align}
    where $N$ is the total number of galaxies in the group,
    $c\bar{z}$ is the mean velocity of the group, and $c\bar{z}_{i}$
    is the velocity of each member galaxy. This variable is only
    computed for galaxy groups with two or more member galaxies. For
    groups with just one member galaxy, we assign a value of '0'
    to \Gdv.
\item \textbf{Velocity dispersion within $\mathbf{r_{med}}$}:
    Similar to $\mathbf{\sigma_{v}}$. We compute a one-dimensional
    velocity dispersion of the galaxies that are within
    $\mathbf{r_{\mathrm{med}}}$ with
    Equation~\ref{eq:group_vel_dispersion}, but only using
    galaxies within the designated radius from the centre of the
    galaxy group.
\item \textbf{Abundance-matched mass, \mgroup}:
    We estimate the total mass of the group via
    \textit{abundance matching}. This method assumes
    a monotonically increasing relationship between the group total
    luminosity, $M_{r,\textrm{tot}}$, and the dark matter halo mass.
    We adopt the \citet{Warren2006} mass function for this purpose.
\item \textbf{Dynamical mass}:
    We follow the prescription from \citet{Girardi1998} for 
    estimating the group dynamical mass, using \Gdv and
    $\mathbf{R_{\perp,rms}}$ as follows
    \begin{align}
        M_{\textrm{dyn}} = A\times \frac{3\pi}{2}\frac{\sigma_v^{2} R_{\perp,rms}}{G},
    \end{align}
    where $G$ is the gravitational constant. \textit{A} is a fudge 
    factor that we use to remove any systematic offset between 
    the dynamical mass estimate and the true halo mass in the cluster 
    mass regime. Based on tests with our mock catalogs, we set this
    fudge factor to a value of `1.04'. With this value of \textit{A},
    the above equation recovers the correct halo mass for a massive 
    halo in the ideal case where the radius and velocity dispersion
    of the halo are known perfectly.
\item \textbf{Distance to closest cluster}:
    Distance to the closest cluster of galaxies that is at
    least a factor of \textit{10} times more massive than the 
    host group of the galaxy in question. Masses are measured
    using halo abundance matching and the distance is in units 
    of \hmpc and is calculated in three-dimensional space.
    If no such cluster of galaxies is to be found, we assign
    a value of '0' to this variable.
\end{enumerate}
This list of features contains spectro-photometric properties of
the galaxies, sizes and velocity dispersions of their groups, two
halo mass estimates (one derived from spectro-photometric properties,
i.e., \HAMs, and one derived from group size and velocity dispersion,
i.e., \DYNs), a group morphological parameter, and a large-scale 
environmental metric. All of these features are expected to contain 
information about halo mass.

\subsection{Machine Learning algorithms}
\label{subsec:ml_algorithms}

Machine learning is an inventive field in computer science,
with a variety of different applications in a number of 
areas. As mentioned in \refsec{sec:intro}, ML algorithms are
able to \textit{learn} non-parametric relationships between
some input \textit{data} and an expected output, without having
to explicitly provide an analytic prescription. In the case of
\textit{supervised} learning, which is the type of ML used
in this paper, a training dataset $(\bX, \by)$ is
provided, and the ML algorithms try to \textit{learn} the
mapping $\bF(\bX \to \by)$ between the set of features, $\bX$,
and the expected output, $\by$. Once the algorithm is trained, 
it is tested on a different `test' dataset in order to quantify 
how well it works. Ultimately, the goal is to apply the algorithm
to an application dataset where $\by$ is not known.

For our study, we test the performance of 3 different
flavours of ML algorithms in order to see which algorithm
can provide us with the best prediction for the halo masses
of galaxies. We use the \textit{Random Forest} an 
\textit{Neural Network} algorithms from the python package 
\sklearn\footnote{\SKurl \label{ft:sklearn}} \citep{Pedregosa2012},
as well as the \XGBoosts algorithm
\footnote{\XGurl \label{ft:xgboost}}.

\subsubsection{Random Forest}
\label{subsub:ml_random_forest}

One of the ML algorithms that we use in this analysis is
\textit{Random Forests} \citep[hereafter \RF;][]{Breiman2001}.
A random forest is an ensemble learning technique that
builds upon a collection of tree-structured classifiers,
also known as \textit{decision trees}. For the purpose
of this analysis, we implement \RF\ for regression rather
than for classification, and decision trees are to be
referred as \textit{regression trees} in this context.
\RF\ makes use of the \textit{bagging} method, in which
it generates \textit{n} samples from the dataset, trains
each sample individually and averages all of the predictions
at the end. For a more comprehensive account of this technique, 
the reader is referred to \citet{Breiman1984}. We implement 
the \sklearn\ version of \RF, \texttt{RandomForestRegressor}, 
with its default settings.

\subsubsection{XGBoost}
\label{subsub:ml_xgboost}

\XGBoost\ \citep{Chen2006} is part of the family of
\textit{boosting} algorithms, which makes use of
the \textit{boosting} method. In Boosting, unlike in
\textit{Bagging}, the algorithm generates $n$
random samples for training with replacement over
weighted data. Each of these regression trees are
referred to as \textit{weak learners}, and they each
get assigned weights based on the accuracy of their
predictions. After these weak learners are
trained, the weighted averages of each of their estimates
are used to compute the final predictions. The combination
of weak learners is referred to as \textit{strong learners}.
For a more in-depth discussion of \XGBoosts and its different
features, the reader is referred to the online documentation
\footref{ft:xgboost}.

\subsubsection{Neural network}
\label{subsub:ml_neural_network}

The last ML algorithm used in this analysis is the simplest 
type of a neural network (\NN), i.e. the 
\textit{Multi-Layer Perceptron} (\MLP). A \MLP\ is a model with 
interconnected information processing units, often referred to as 
\textit{neurons}, that learns the mapping $\bF(\bX \to \by)$ 
given a training set $(\bX, \by)$, with $\bX$ being the input 
features and $\by$ the target elements to predict. We implement 
the \sklearn\ version of a 3-layer \MLP\ with each layer 
containing 100 \textit{neurons}. We refer the user to the \sklearn\ 
documentation\footref{ft:sklearn} for a more comprehensive account 
of this method.

%% file: 04_training_testing.tex

\section{Training and Testing ML algorithms}
\label{sec:training_testing}

In this section, we present results from the training
and testing of the three ML algorithms for predicting
the halo masses of galaxies in SDSS DR7. Moreover, we 
compare these predictions to the more traditional
estimates from halo abundance matching (\HAM) and
dynamical mass measurements (\DYN). In \refsec{subsec:feat_selection},
we present the set of features that contribute the most to
the overall prediction of halo mass in order to reduce the 
dimensionality of our feature space in further training.
In \refsec{subsec:fixed_hod_dv}, we present results from the
training and testing phases of each of the three ML
algorithms using our synthetic catalogues of the Universe. The
mock catalogues used in the training and testing phases are
built using the same HOD model and thus represent the overly
optimistic scenario in which the training data perfectly 
represents the real universe. Results in this section thus
serve as a proof of concept that ML is a feasible method of 
determining the halo masses of galaxies.
We explore the more realistic case that the training data
is drawn from a different underlying model than the real universe
in \refsec{sec:universality}.

\subsection{Feature Selection}
\label{subsec:feat_selection}

\begin{figure}
    \centering
    \includegraphics[width=\figwidth\textwidth]{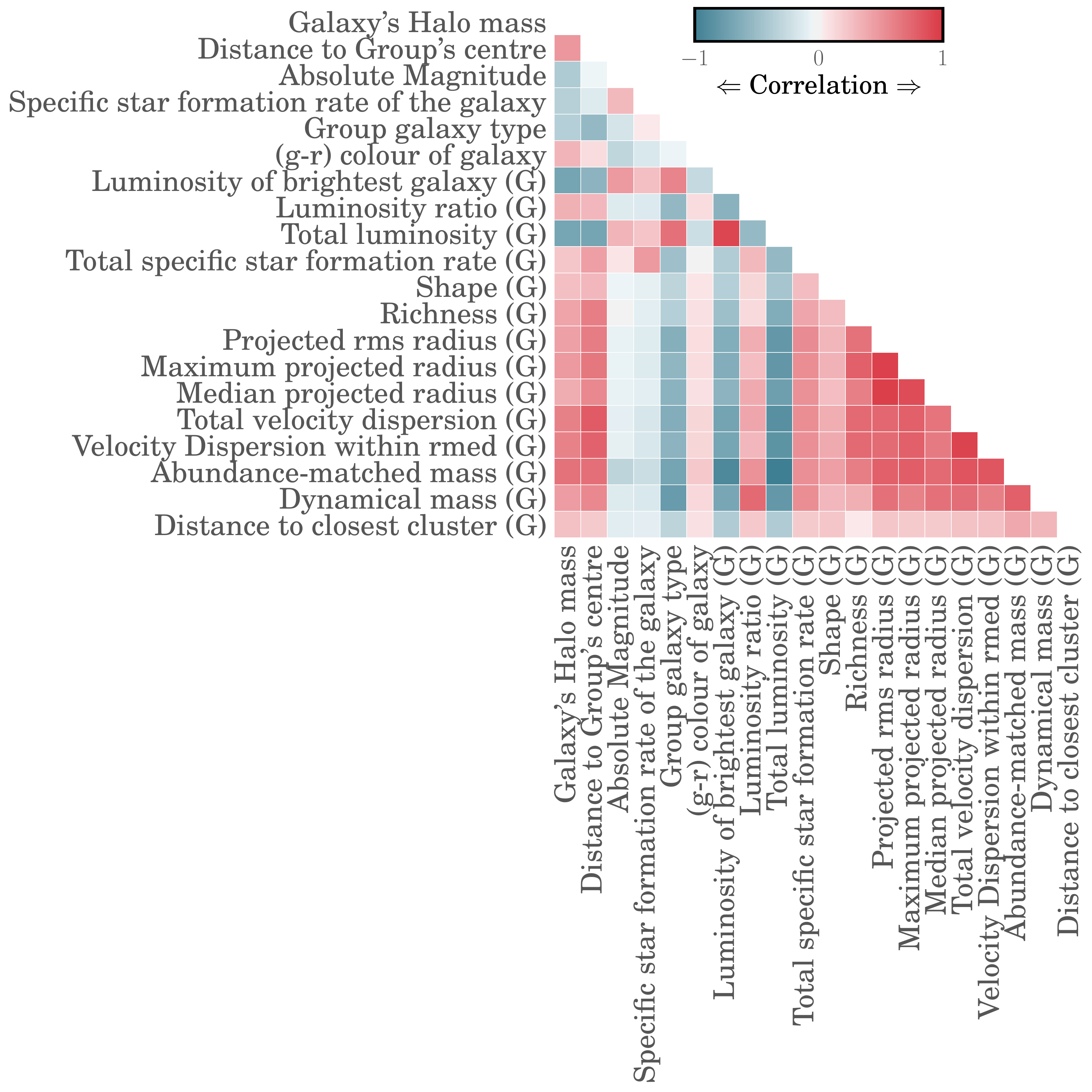}
    \caption{
    Correlation matrix of the galaxy- and group-related features
    presented in \refsec{subsec:galprop_features}, computed using
    our mock galaxy catalogues. The figure shows how correlated the 
    features are with each other, with red and blue shadings 
    corresponding to positive and negative correlations, respectively. 
    Additionally, the first column displays the degree of correlation 
    of each feature with halo mass, which is the quantity we wish
    to predict. This figure conveys the point that the mass of the dark 
    matter halo is strongly correlated with almost all of the features 
    that we consider for training the different ML algorithms.
    }
    \label{fig:corr_matrix_features}
\end{figure}

In \S~\ref{subsec:galprop_features} we presented a list of 19 
properties of galaxies and their groups that may contain useful
information about halo mass. In this section we analyse the 
predictive power of these features in order to eliminate ones
that are not as useful and thus reduce the overall number of 
features that we will use as inputs to the ML algorithms. 
This is conventionally referred to as \textit{feature selection}, 
and it plays an important role into the training process of a ML 
algorithm. Reducing the dimensionality of the feature space is
desirable because it reduces the computational cost of ML 
algorithms and can also improve their predictive performance.

Before we determine the importance of each feature for the
prediction of halo mass, we first explore the amount of correlation
among the different features from \refsec{subsec:galprop_features}.
Figure~\ref{fig:corr_matrix_features} presents the correlation
matrix of these 19 features as measured from our mock galaxy 
catalogues. The matrix shows the correlation coefficient between 
each pair of features, with red and blue shadings corresponding to 
positive or negative correlation, respectively. The matrix also 
includes halo mass in the first column and thus reveals how much 
each feature is correlated with the quantity we are trying to 
predict. Figure~\ref{fig:corr_matrix_features} shows that almost
all 19 of our features exhibit correlations with halo mass. 
Additionally, many of the features are highly correlated with each 
other, as expected, and are thus unlikely to contain independent 
information about halo mass.

To quantify the importance of each feature for the purpose of feature
selection, we use the native feature importance calculation within
the \RFs and \XGBoosts algorithms (the \NNs algorithm does not compute 
such a statistic). In general, these algorithms estimate the importance 
of a feature by calculating how much it is used to make key decisions
with their decision trees. Each feature gets an importance score
allowing us to compare them to each other and rank them.
Though later on we will split our 10 mock catalogues into training and 
testing subsets, for the purpose of feature selection we use them all to 
train the \RFs and \XGBoosts algorithms. Each algorithm then produces a 
ranked list of the 19 features in order of their importance, as 
discussed above. Though the two algorithms differ in their detailed
ranking of features, they are generally consistent and are almost in perfect
agreement on which features land in the top nine (out of 19). The remaining 
set of features do not contribute much to the overall prediction of halo 
mass and so we focus on these nine features moving forward.

\begin{figure}
    \centering
    \includegraphics[width=\figwidth\textwidth]{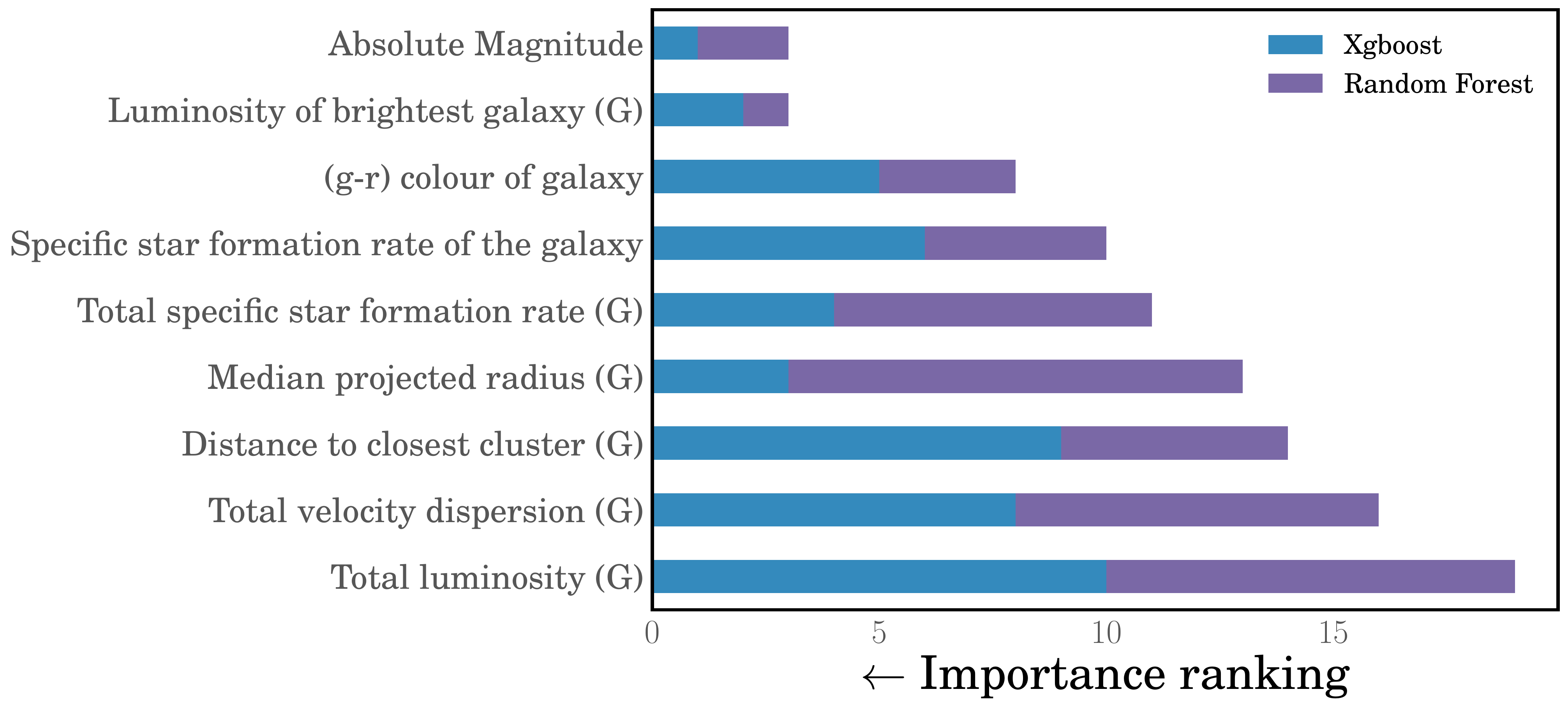}
    \caption{
    Feature importance for the top nine features used when predicting 
    the mass of a galaxy's host dark matter halo, as calculated by
    the \XGBoosts (blue bars) and \RFs (purple bars) ML algorithms.
    The length of each bar indicates its importance rank, with shorter 
    bars corresponding to more important features.
    }
    \label{fig:feat_importance_ranking}
\end{figure}

Figure~\ref{fig:feat_importance_ranking} shows the feature importance
ranks for these top nine features for both the \XGBoosts and \RFs 
algorithms. In the case of each feature, the length of the blue and 
purple bar indicates its importance rank as calculated by \XGBoosts and 
\RFs, respectively, with shorter bars corresponding to more important 
features. We estimate the overall importance of each feature by adding 
its two ranks (the combined length of the blue and purple bars) and we 
order the features in Figure~\ref{fig:feat_importance_ranking} according 
to this overall score. The figure shows that the luminosity of the galaxy 
itself and the luminosity of the brightest galaxy in the galaxy's group are 
the overall most useful features in predicting halo mass, while the total 
group luminosity is the least useful from this set of top nine features.

We select these top nine features that contribute the most to the 
prediction of halo mass as our final set of features moving forward.
\paragraph*{Final set of features}
\begin{enumerate}[label=\arabic*, leftmargin=1\parindent, labelsep=0.5\parindent]
    \itemsep0em
    \item Galaxy's \rband\ absolute magnitude
    \item Luminosity of the \textit{brightest} galaxy in the group
    \item Galaxy's \gr colour
    \item Galaxy's specific star formation rate
    \item Group's total specific star formation rate
    \item Group's median projected radius
    \item Distance to the closest cluster
    \item Group's total velocity dispersion
    \item Group's total \rbands absolute magnitude

\end{enumerate}
For the rest of the analysis in this paper, we will exclusively use
this set of features to train the various ML algorithms and evaluate 
their performance at correctly predicting halo masses.


\subsection{Training and Testing}
\label{subsec:fixed_hod_dv}

Now that we have a final list of nine input features, we can proceed
to the training and testing of the ML algorithms. We start with our
set of 10 mock galaxy catalogues, each of which has the same volume
and approximate number density as the \MD{19} sample. Combined, these 
catalogues contain a total of 758,528 mock galaxies. For each galaxy
we have values for the nine input features as well as the target 
halo mass. We also have the traditional \HAMs and \DYNs mass 
measurements to compare against.

We split the mock data into \textit{training} and \textit{testing}
sets. The training set consists of 8 of the 10 catalogues, while the
testing set consists of the remaining 2. We will use the testing set to
evaluate how well the trained algorithms perform. It is important to 
perform this evaluation on an independent set of data from the training
set in order to guard against the problem of over-fitting. Sometimes ML 
analyses also use a third, \textit{validation}, dataset for the purpose 
of tuning the hyper-parameters of a given ML algorithm. However, in this 
paper we choose to adopt the default values of hyper-parameters and thus 
we do not need to add a validation step to our workflow.

After training the three ML algorithms to predict the dark matter halo 
masses of mock galaxies in the training set, we apply these trained 
algorithms to the testing data and get a list of predicted masses, \mpred,
for these galaxies. We then compare these predictions against the true 
halo masses, \mtrue, and compute the fractional difference between their
logarithmic values as
\begin{align}\label{eq:frac_diff}
    \fracdm &= 100\times\left[\frac{\log \mpredm}{\log \mtruem} - 1 \right] .
\end{align}
Each galaxy in the testing set gets three values of $\fracdm$ (one for 
each ML algorithm), which are essentially the fractional errors in the ML 
predictions. Note that these are errors in the \textit{logarithm} of halo
mass. A value of $\fracdm=$5\% thus corresponds to a fractional error in 
mass of $\sim250-400$\% for the mass range we consider here. For comparison, 
we also calculate $\fracdm$ using the \HAMs and \DYNs masses in place of 
\mpred. This will allow us to examine how well the ML algorithms perform 
relative to traditional methods for estimating halo mass.
\begin{figure}
    \centering
    \includegraphics[width=\figwidth\textwidth]{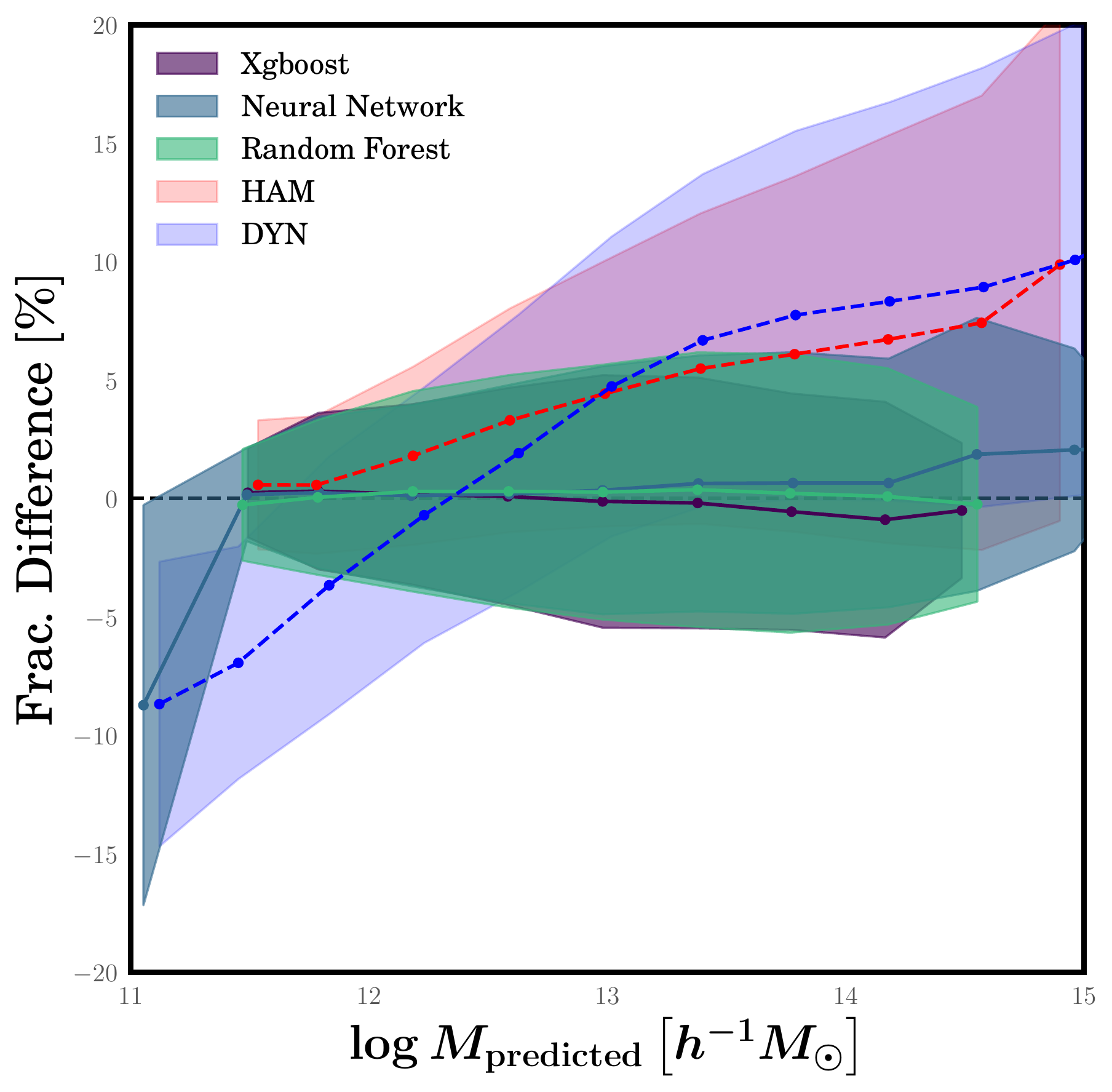}
    \caption{
    Fractional difference between \textit{predicted} and
    \textit{true} logarithmic halo mass for galaxies, as a 
    function of predicted halo mass, for different methods of 
    estimating the dark matter halo mass of a galaxy. Results 
    are shown for a testing set of mock galaxies, for which their 
    true masses are known. The solid, coloured lines correspond 
    to the mean fractional difference of each method, while the 
    shaded regions represent the \sig{1} ranges. This figure 
    compares the predictions of halo mass made by the three 
    different ML algorithms to the estimates from conventional 
    methods, i.e. halo abundance matching (\HAM) and dynamical 
    mass estimates (\DYN).
    }
    \label{fig:hod_dv_frac_diff_mass}
\end{figure}

Figure~\ref{fig:hod_dv_frac_diff_mass} presents results for \fracd, as a 
function of predicted mass, for different methods of estimating the halo 
masses of galaxies. The solid, coloured lines correspond to the mean 
fractional difference of galaxies in bins of \mpred, while the shaded
regions represent the \sig{1} ranges of \fracd. We show predictions
made by the \XGBoost, \RF, and \NNs algorithms, and compare these to the 
mass estimates obtained from \HAM\ and \DYN. 
Figure~\ref{fig:hod_dv_frac_diff_mass} shows promising results, in
that all three ML algorithms are performing significantly better at 
predicting the mass of a galaxy's host halo than either \HAMs and \DYN.
Specifically, \HAMs yields halo masses that are unbiased on average at low
masses and have a \sig{1} error of $\sim3\%$, but it systematically 
overestimates masses on average at high masses, reaching a systematic
error as high as $\fracdm\sim10\%$ in the cluster regime. Moreover, the 
scatter grows to $\sim10\%$ in this regime as well. \DYNs exhibits even 
worse performance since it has similar poor performance for large masses, 
but also does badly at low masses, systematically underestimating masses on 
average as much as $\fracdm\sim10\%$. In contrast, the three ML algorithms 
yield predicted masses that are unbiased on average at all masses and have 
a \sig{1} scatter in \fracds of $\sim3-5\%$.

To understand the poor performance of the \HAMs and \DYNs methods, it is
important to consider that we are not evaluating the ability of these 
methods to correctly estimate the halo masses of galaxy \textit{groups}, 
but rather of individual \textit{galaxies}. Grouping errors made by the 
group-finding algorithm can thus cause catastrophic errors in the halo 
masses of galaxies that have been incorrectly grouped. For example, if the 
group-finder incorrectly merges together a few galaxies that live in small 
haloes with the galaxies of a large halo to yield a single massive galaxy 
group, both \HAMs and \DYNs will estimate a large halo mass for this group 
and, thus, for all its members. The error in this estimate will be small 
for the galaxies that actually belong to the large halo, but will be 
enormous for the galaxies that were mistakenly grouped. It is these 
catastrophic errors that drive both methods to overestimate the masses of 
galaxies on average in the high mass regime in 
Figure~\ref{fig:hod_dv_frac_diff_mass}. At low masses, where most 
galaxies live in $N=1$ groups, \HAMs does a good job at recovering the mass 
because galaxy luminosity correlates strongly with mass. \DYN, however, 
does poorly because dynamical measurements are very unreliable for systems 
with a small number of galaxies. The ML algorithms have the advantage that 
they use additional information that can help fix some of the problems 
caused by grouping errors. In the example above, the colours of incorrectly 
grouped galaxies are likely different from those of actual satellite 
galaxies in massive halos and the ML algorithms exploit this to distinguish 
between the two. An exciting possibility that arises from this is that the 
halo masses predicted by ML could be used to improve the group-finding 
itself since galaxies whose predicted masses are much smaller than the 
groups they've been assigned to could be removed from them. We return to 
this point in the final section.

Another way to quantify the effectiveness of these algorithms at predicting 
halo masses is to determine the percentile discrepancy between the true and 
predicted halo masses across a big range of \mpreds masses. To compute this 
statistic, we first determine the absolute value of the log-difference 
between predicted and true halo mass, and rank-order them from smallest to 
largest. We then determine the discrepancy that corresponds to the 68\% of 
galaxies that are best predicted. This statistic is given by the following 
equation:
\begin{align}\label{eq:mass_discrepancy}
    \deltalogmm &= \calP_{68}\left(\ \left| \mathrm{log} M_{\mathrm{pred}} - \mathrm{log} M_{\mathrm{true}} \right|\ \right).
\end{align}
In other words, 68\% of galaxies have their masses predicted with an error 
less than \deltalogm. We split the test sample into a \textit{low-mass} and 
\textit{high-mass} galaxy sample. Galaxies with $\log \mpredm \leq 12.5$ 
are assigned to the \textit{low-mass} sample, while those with 
$\log \mpredm > 12.5$ are assigned to the \textit{high-mass} sample. For 
each sample, we compute \deltalogms for each ML algorithm, and compare them 
to those for \HAMs and \DYN. This statistic shows how well each method is 
at estimating the halo masses in these two mass regimes.

\begin{figure}
    \includegraphics[width=\figwidth\textwidth]{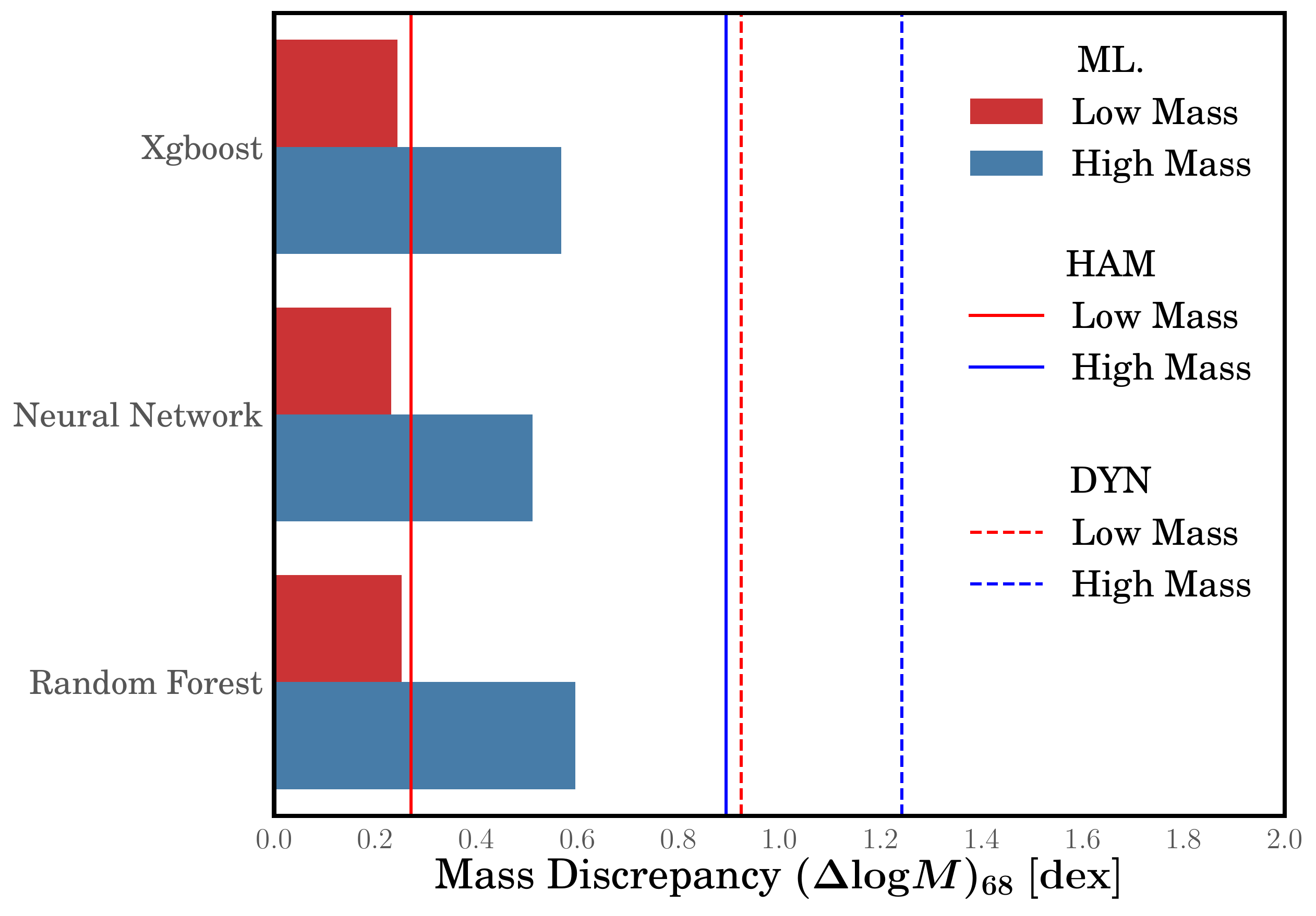}
    \caption{
    Mass discrepancies, \deltalogms, for the three ML algorithms, as 
    compared to those of \HAMs and \DYNs methods, when splitting the galaxy
    sample into \textit{low-mass} and \textit{high-mass} samples.
    The quantity \deltalogms is the 68\% prediction error in the log of 
    halo mass, meaning that 68\% of galaxies are predicted better than this.
    The horizontal bars show this typical error for the ML algorithms while 
    the solid and dashed vertical lines correspond to the \HAMs and \DYNs
    methods, respectively. In all cases, results for galaxies with 
    $\log \mpredm \leq 12.5$ are shown in red, while results for galaxies
    with $\log \mpredm > 12.5$ are shown in blue. The three ML algorithms
    exhibit similar performance and are significantly better than 
    traditional methods, especially in the high mass regime.
    }
    \label{fig:hod_dv_sigma_err_general}
\end{figure}

Figure~\ref{fig:hod_dv_sigma_err_general} presents the results for the
typical mass error \deltalogm. Horizontal bars show values for the three
ML algorithms, while solid and dashed vertical lines show results for the
\HAMs and \DYNs methods, respectively, for comparison. In all cases, 
results for galaxies with low predicted masses are shown in red, while 
results for galaxies with high predicted masses are shown in blue.
Figure~\ref{fig:hod_dv_sigma_err_general} shows clearly that the three ML
algorithms exhibit similar performance and they significantly outperform 
traditional methods in most cases. \HAMs does well at low masses, but at 
high masses its error is $\sim 50-60\%$ larger than ML methods. \DYNs does 
poorly in both mass regimes, with a typical error that is $2-4$ times larger
than that for ML methods. More specifically, \HAMs is able to estimate
halo masses to within a $\deltalogmm \approx 0.27$ dex and 
$\deltalogmm \approx 0.90$ dex for the low-mass and high-mass regimes,
respectively. On the other hand, \DYNs can only recover halo masses to 
within $\deltalogmm \approx 0.92$ dex and $\deltalogmm \approx 1.25$ dex 
for the low-mass and high-mass regimes, respectively. The corresponding 
errors for the \XGBoost, \RF, and \NNs ML algorithms range from, 
$\deltalogmm \approx 0.23-0.25$ dex and $\deltalogmm \approx 0.51-0.60$ dex 
for the low-mass and high-mass samples, respectively. 

In summary, we find that we are able to obtain better mass estimates for a
galaxy's host halo by using ML methods in place of the more traditional mass
estimators, such as \HAMs or \DYNs. This statement is true regardless of 
predicted mass, \mpred. However, so far this statement only holds for the 
case in which the training and testing samples share the same underlying 
model that connects galaxies to dark matter halos. This is not likely to be 
true when we apply the trained models to real SDSS data. We address this 
issue in the next section.

%% file: 05_universality_in_model.tex

\section{Are Mock-Trained Models Universally Applicable?}
\label{sec:universality}

The results shown in \refsec{subsec:fixed_hod_dv} support the notion
that we can obtain better halo mass estimates for galaxies by employing 
ML algorithms instead of the traditional \HAMs or \DYNs methods. We 
evaluated the performance of the ML algorithms using a testing set of mock 
galaxy catalogues that are independent from the set that we used to train 
the models. In this context, ``independent'' means that they are 
constructed from cosmological N-body simulations that are independent 
realisations of the density field (i.e., have initial conditions with 
different random phases). However, the testing catalogues adopt the same 
prescription for populating dark matter halos with galaxies and assigning 
them observed properties like luminosity and colour. A better approach 
would be to test the ML algorithms using catalogues that were built with 
different such prescriptions, since the real universe is unlikely to 
perfectly conform to the assumptions made in the training phase. In this 
section, we test the impact of these assumptions in order to assess whether 
mock-trained models can be applied to the real universe.


\subsection{Varying HOD models}
\label{subsec:varying_hod}

The first step we make to build mock galaxy catalogues from a dark matter 
halo distribution is to populate the halos using a HOD model. This model 
specifies the number of central and satellite galaxies that are placed in 
each halo. The model is flexible and has five free parameters. We use the 
best-fit parameter values of \citet{Sinha2018}, which ensure that the 
number density, clustering, and group statistics of our catalogues match 
those observed in the SDSS. This is the \textit{fiducial} HOD model that we 
used to train and test our models in \S~\ref{sec:training_testing}. To test 
how sensitive our results are to the HOD model of the testing sets, we now 
produce different versions of our two synthetic testing catalogues, each 
with different values for the five HOD parameters. We select the parameter 
sets from the \citet{Sinha2018} MCMC chain so that the resulting mock 
catalogues are still consistent with SDSS observations. We then run the 
previously trained ML algorithms on these new test mock catalogues to 
investigate how much performance we lose from modifying the HOD model in 
the testing phase.

\begin{figure}
    \centering
    \includegraphics[width=\figwidth\textwidth]{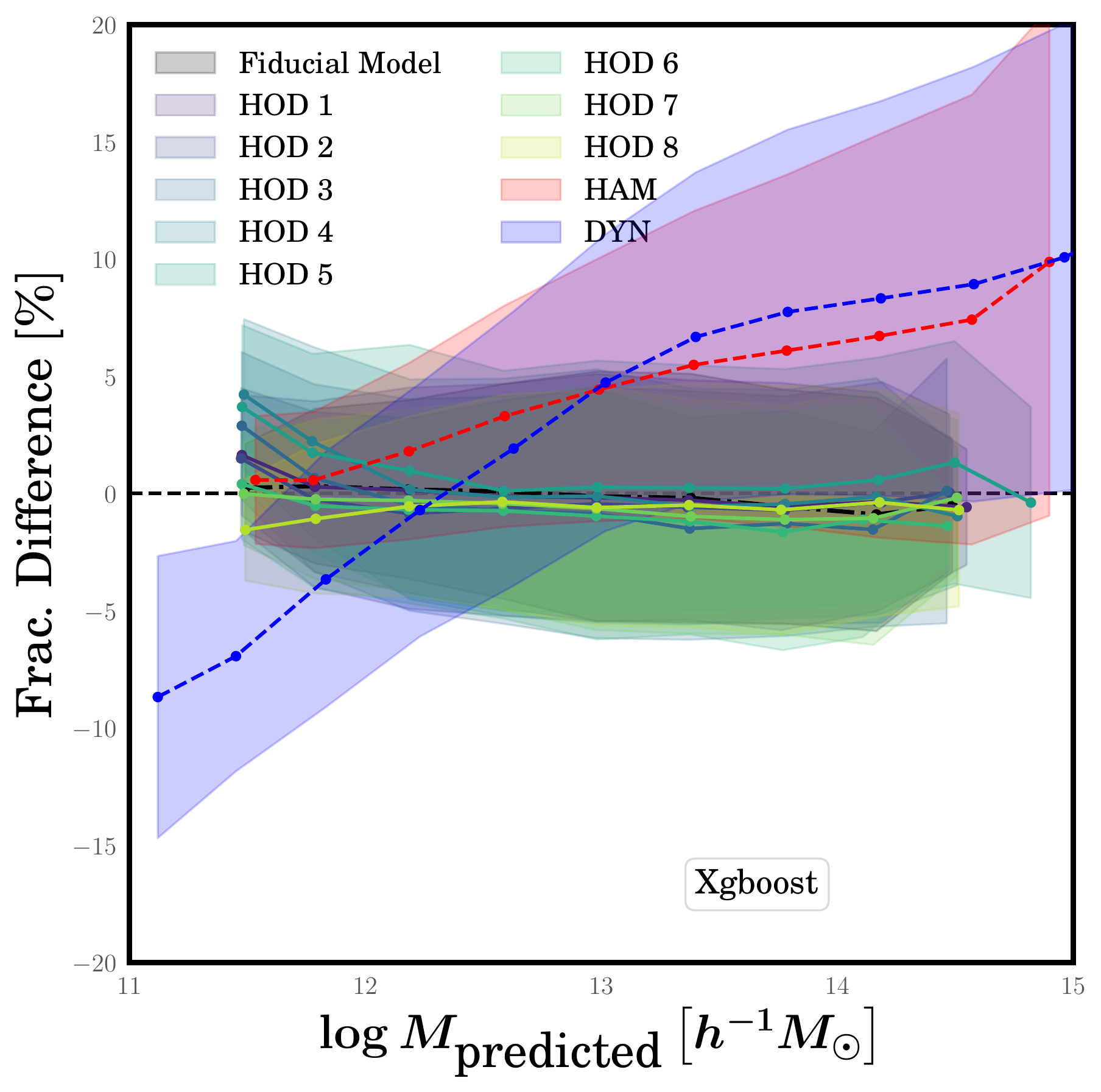}
    \caption{
    Fractional difference between \textit{predicted} and
    \textit{true} logarithmic halo mass for galaxies, as a 
    function of predicted halo mass, for a variety of testing data
    sets that were constructed using different halo occupation
    models than what was used in the training phase. All models
    shown use the \XGBoosts ML algorithm. Lines and shaded regions
    have the same meaning as in Fig.~\ref{fig:hod_dv_frac_diff_mass}.
    Results for \HAMs and \DYNs are also shown for comparison (for the
    fiducial model case).
    }
    \label{fig:hod_frac_diff_mass}
\end{figure}

Figure~\ref{fig:hod_frac_diff_mass} shows the fractional difference between 
predicted and true halo mass, \fracd, for these new test sets. The figure 
is similar to Figure~\ref{fig:hod_dv_frac_diff_mass}, except that it only 
shows results for the \XGBoosts algorithm and it focuses on the different 
HOD models instead. Also shown are the \HAMs and \DYNs results for 
comparison, which are applied to the fiducial test catalogues. We have also 
done the same tests using the \RFs algorithm and obtained similar results.
Figure~\ref{fig:hod_frac_diff_mass} reveals that the performance of the ML 
algorithm degrades significantly at low masses when it is applied to 
testing catalogues with different HOD models. For predicted masses larger 
than $\gtrapprox 10^{12} \msun$ the effect is negligible and ML clearly 
outperforms the \HAMs and \DYNs methods just as it did when tested on the 
fiducial model. However, for $\mpredm \lessapprox 10^{12} \msun$, the mean 
\fracds is significantly biased for some of the HOD models, reaching values 
as high as 4\%. 

To understand why the ML algorithms degrade at low \mpred, we take a close
look at the HOD parameters of our models to see if there is a trend that
explains why some models result in high \fracds while others do not.
We find a very strong correlation between \fracds and 
$\sigma_{\mathrm{log}M}$, the scatter in halo mass at the luminosity limit
of the sample. Test catalogues with high values of this scatter receive 
predicted masses that are systematically overestimated when trained using 
the fiducial model. The fiducial model adopts a value of 
$\sigma_{\mathrm{log}M}=0.14$ \citep{Sinha2018}, while the most extreme 
HOD models we test have values of $0.5-0.9$. Increasing the scatter this 
much is equivalent to removing some central galaxies from larger halos and 
placing them in lower mass halos. However, their observed properties 
(e.g., luminosity and colour) don't change much because they are assigned in
a way that perfectly recovers the observed distributions in the SDSS. For 
example, in our mock catalogues the faintest $r$-band absolute magnitudes for 
mock galaxies are always equal to $-19$ regardless of their halo mass, since 
that is the luminosity limit of our SDSS sample. As a result, ML algorithms 
trained on a catalogue where these faintest galaxies live in more massive 
haloes, but applied to a catalogue where they live in less massive halos, will 
learn an incorrect mapping between luminosity and halo mass and thus predict 
masses that are too high. 

Figure~\ref{fig:hod_frac_diff_mass} suggests that in the low mass regime, the 
\HAMs method can yield more reliable halo masses than the ML algorithms. 
However, this is not the case. The \HAMs result shown is only for the fiducial 
model and performs well at low mass. However, the \HAMs method applied to the 
other HOD models exhibits even worse performance than the ML algorithms. The 
reason for this is that catalogues built assuming a high 
$\sigma_{\mathrm{log}M}$ have their lowest luminosity galaxies living in lower
mass haloes than they do in catalogues with a smaller scatter, but their number 
density is not correspondingly higher because not all haloes down to this mass 
are occupied. Since the \HAMs method uses abundances to assign mass, it will 
overpredict these galaxies' masses. So even though ML does poorly when applied to
high $\sigma_{\mathrm{log}M}$ datasets, it still outperforms \HAM. Another 
thing to consider is that the ML algorithms only perform poorly when applied to 
very large values of $\sigma_{\mathrm{log}M}=0.5-0.9$, which are likely 
inconsistent with observed data. The true amount of this scatter in the real 
universe is most likely close to $\sim0.2$ where our trained ML algorithms 
perform quite well.

\subsection{Varying Satellite Galaxy Velocity bias}
\label{subsec:varying_dv}

In the previous section, we demonstrated the effect of varying the HOD
parameters that control the number of central and satellite galaxies that
occupy haloes as a function of mass. Now we investigate varying how we 
place these galaxies in their haloes when we construct test mock catalogues.
Specifically, we study the effect of adding velocity bias to our mocks.
In the fiducial model, satellite galaxies are assigned the positions and 
velocities of randomly selected dark matter particles within their haloes.
However, it is possible that satellite galaxies have kinematics that are 
either hotter or colder than the underlying dark matter
\citep[e.g.,][]{Guo2015}. This is referred to as \textit{velocity bias}.
We parameterise this bias as the ratio between the velocity dispersion
of satellite galaxies, $\sigma_{v,\mathrm{sat}}$, within a halo and the 
velocity dispersion of dark matter, $\sigma_{v,\mathrm{dm}}$,
\begin{align}\label{eq:vel_bias}
    \sigma_{v,\mathrm{sat}} &= \fvbm\times\sigma_{v,\mathrm{dm}} ,
\end{align}
where \fvbs is the velocity bias parameter, and we explore models with
values between \fvbs=0.9 and 1.1. We implement velocity bias into our mock 
catalogues simply by scaling satellite galaxies' assigned velocities by 
\fvbs. Velocity bias is important in this ML context because it directly 
affects dynamical measurements of group mass. A test mock catalogue with 
velocity bias will have a different relationship between group velocity 
dispersion and halo mass, which could cause errors in the predicted mass 
since velocity dispersion is a feature used by the ML algorithms. In 
addition, velocity bias will change the size of small-scale redshift 
distortions in groups, which can affect grouping errors.

\begin{figure}
    \centering
    \includegraphics[width=\figwidth\textwidth]{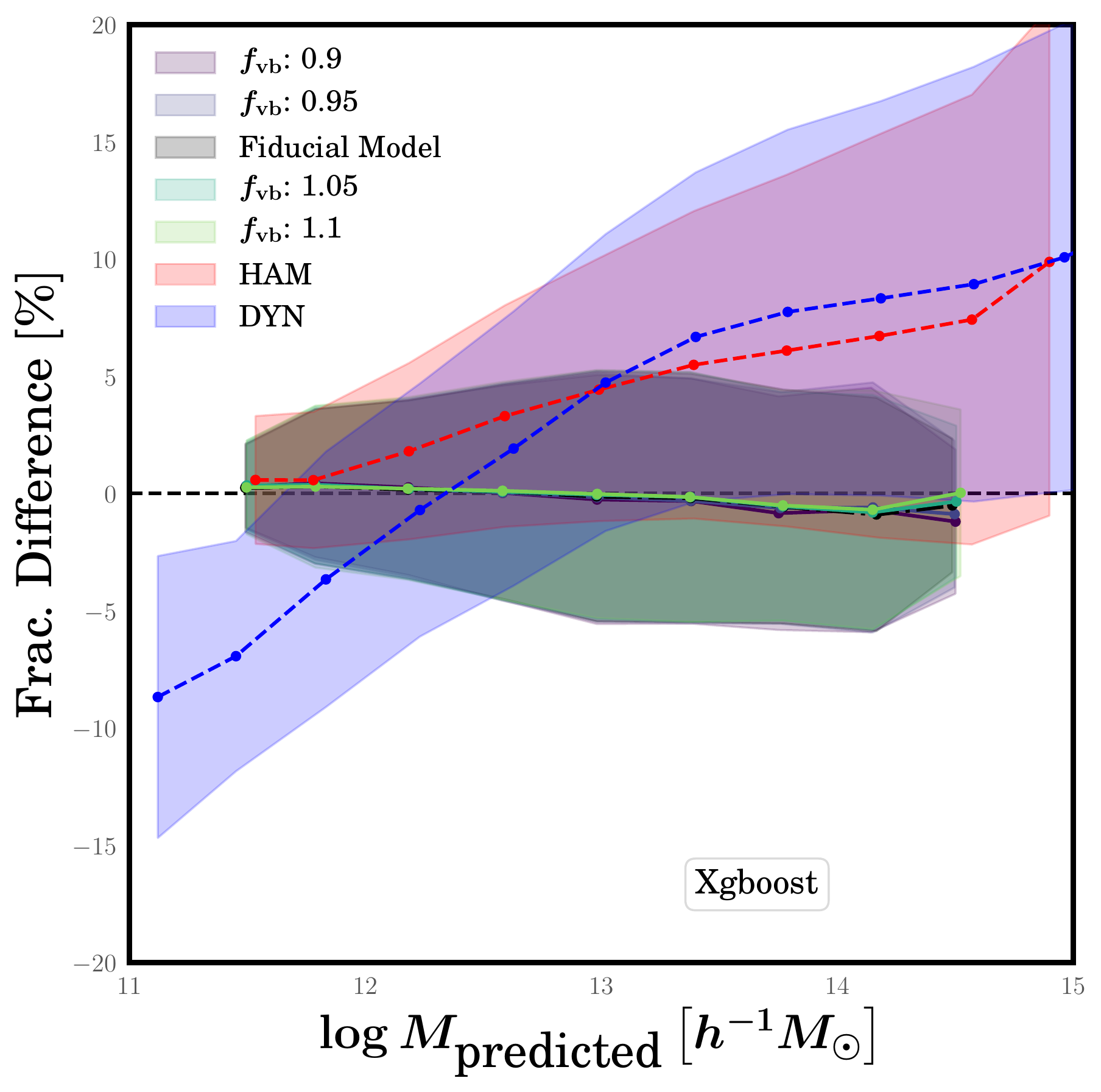}
    \caption{
    Similar to Fig.~\ref{fig:hod_frac_diff_mass}, except that the
    various testing data-sets now share the same set of halo occupation 
    parameters as the training data, but cover a wide range of different 
    values for satellite galaxy velocity bias, \fvb.
    }
    \label{fig:dv_frac_diff_mass}
\end{figure}

To probe the effect of velocity bias on the performance of the ML
algorithms, we construct a few sets of the two testing mock catalogues,
each time adopting the fiducial HOD model, but adding an amount of velocity
bias between \fvbs=0.9 and 1.1. We then apply our previously trained ML 
algorithms to these new test sets. Figure~\ref{fig:dv_frac_diff_mass} shows
the fractional difference \fracds for these test cases compared, as always,
to the \HAMs and \DYNs methods. We only show results for the \XGBoosts
algorithm, but the other algorithms exhibit similar behaviour. The figure 
shows clearly that the performance of ML is almost entirely unaffected by 
velocity bias. This is to say that, regardless of the choice of \fvbs in 
the testing catalogues, the predictions of halo mass made by ML algorithms 
that were trained on the fiducial model are not biased by this choice of 
parameters.

\subsection{Varying the Luminosity-Mass relation}
\label{subsec:varying_scatter}

Having explored the impact of training ML models on data sets that assume
incorrect relationships between the numbers and velocities of galaxies
with halo mass, we now turn to assumptions about the mass-luminosity 
relation. This is potentially important since our feature selection
procedure showed that a galaxy's luminosity and the luminosity of the
brightest galaxy in its group are the two most important features for
predicting halo mass. In our mock catalogues, we assign luminosities to
galaxies using the \textit{Conditional Luminosity Function} (CLF) 
formalism of \citet{Cacciato2009}. Within the CLF model, the main parameter
that controls the strength of the correlation between the mass of a halo 
and the luminosity of its central galaxy is \siglogl, which is the scatter
in the log of luminosity of central galaxies at fixed halo mass.\footnote{In 
\citet{Cacciato2009} this parameter was called $\sigma_c$.}
In the fiducial model that we used to train the ML algorithms, the value of 
this scatter is \siglogl=0.142. To investigate the effect of applying 
the algorithms to data with different correlation between halo mass and 
luminosity, we construct sets of our two test catalogues that assume 
different values of \siglogl, ranging from 0.1 to 0.3.

\begin{figure}
    \centering
    \includegraphics[width=\figwidth\textwidth]{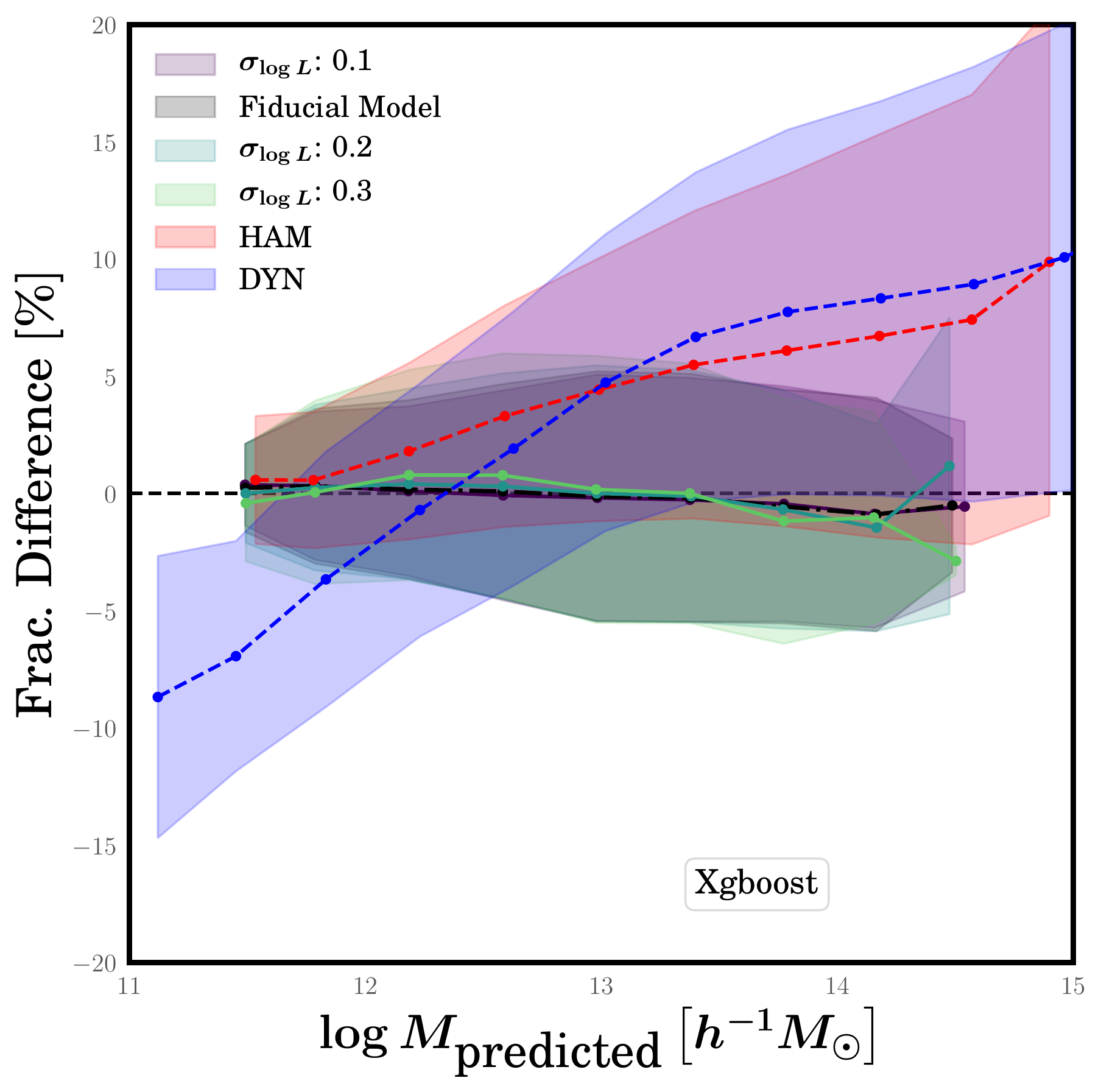}
    \caption{
    Similar to Figs.~\ref{fig:hod_frac_diff_mass}
    and~\ref{fig:dv_frac_diff_mass}, except that the various testing data-sets now share the same set of halo occupation parameters as the training data and have no velocity bias, but cover a range of different values for the assumed scatter in the luminosity-mass relation for central galaxies, \siglogl.
    }
    \label{fig:sigL_frac_diff_mass}
\end{figure}

Figure~\ref{fig:sigL_frac_diff_mass} shows the fractional difference 
\fracds for these test cases. As before, we only show results for the 
\XGBoosts algorithm and we include the results for \HAMs and \DYNs for
comparison. The figure shows that the performance of ML algorithms is
not affected much by the assumed value of \siglogl. This is reassuring and 
implies that our halo mass predictions are not sensitive to the detailed 
form of the mass-luminosity relation.

%% file: 06_SDSS.tex

\section{Application to SDSS Galaxies}
\label{sec:applying_sdss}

In \refsec{sec:training_testing} and \refsec{sec:universality},
we showed how machine learning algorithms, such as \XGBoost,
\RF, and \NN, can be used to predict the mass of a galaxy's
host halo with a higher accuracy on average than more conventional
mass estimators, such as \HAMs and \DYN. The next logical step is to
choose the best of these algorithms and apply the trained 
model to \textit{real} observed data. All three ML algorithms that 
we have explored perform very similarly so we choose \XGBoosts to be 
our algorithm of choice because it is faster than \RFs and \NN.
We apply the \XGBoosts model that we trained and tested on mock 
catalogues to the \MD{19} catalogue, using the nine
features described in \refsec{subsec:feat_selection} as inputs
to the model. The model outputs a predicted halo mass, \mpred, for
each SDSS galaxy. We produce a final catalogue that includes 
the set of nine features for each galaxy in the sample, our value 
for \mpred, and the \HAMs and \DYNs group mass estimates. The 
catalogue is available for download.
\footnote{\catsurl\label{footn:catsurl}}

\begin{figure}
    \centering
    \includegraphics[width=\figwidth\textwidth]{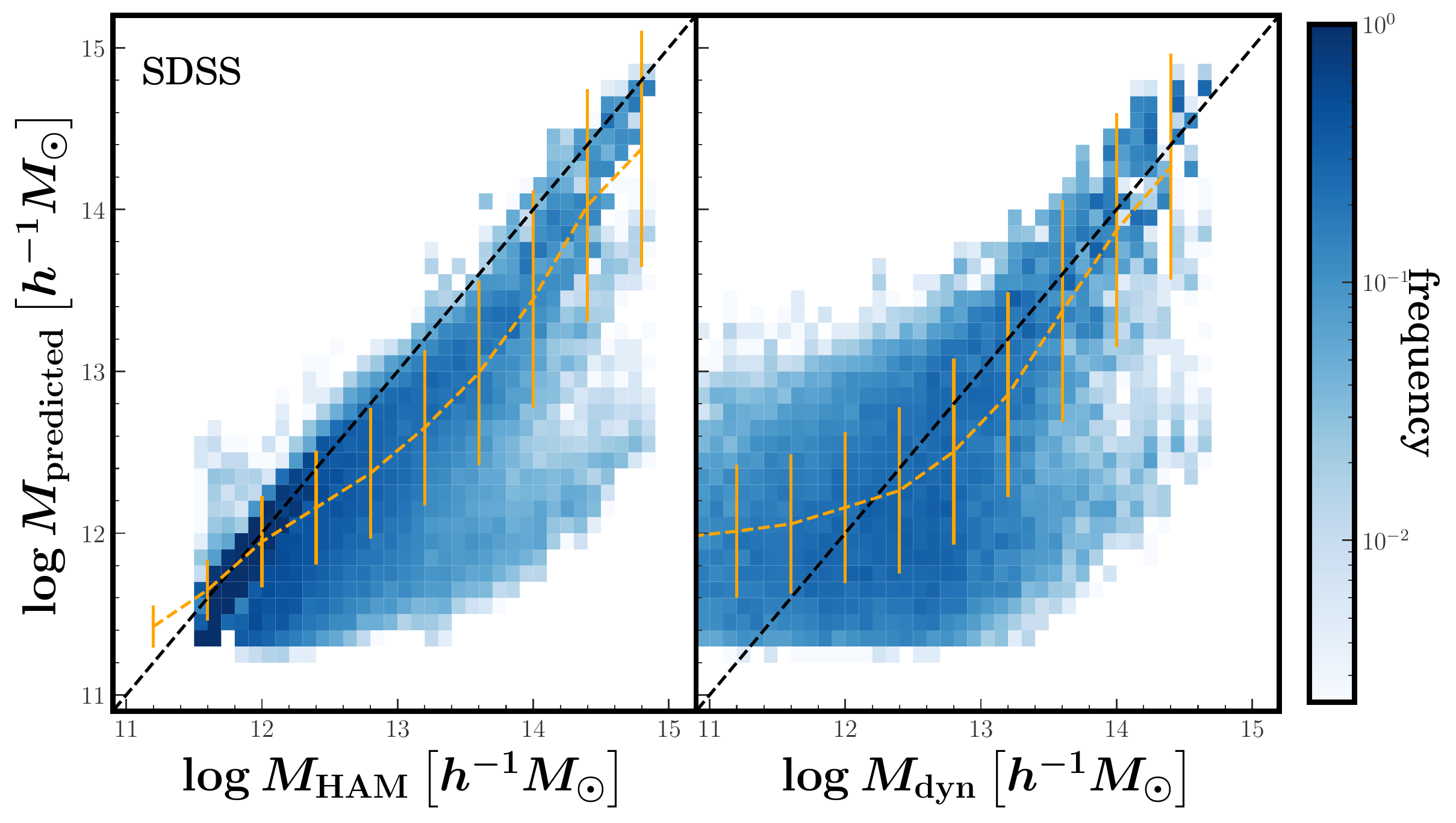}
    \caption{
    Galaxy halo masses for SDSS galaxies predicted by ML compared to 
    traditional methods. The y-axis shows galaxy mass predictions from the 
    \XGBoosts algorithm that was trained on mock catalogues. The x-axis 
    shows mass estimates for galaxies through \HAMs (\textit{left panel})
    and \DYNs (\textit{right panel}). The blue shading shows the frequency
    of galaxies in two-dimensional bins, where the number of galaxies in
    each bin is normalised by the value for the bin containing the most 
    galaxies. Yellow solid lines and errorbars correspond to the mean 
    and standard deviation of \mpreds in bins of \Mhams or \Mdyn. The
    dashed black lines show the one-to-one relation between mass
    estimates.
    }
    \label{fig:data_model_mass_comparisons}
\end{figure}

Figure~\ref{fig:data_model_mass_comparisons} shows the relationship
between \mpreds for SDSS galaxies and the masses from the \HAMs and
\DYNs methods. The figure shows the two-dimensional histogram (blue
shaded pixels) as well as the mean and standard deviation of \mpreds 
in bins of \Mhams and \Mdyns (yellow lines and error bars). In the 
case of \HAM, Figure~\ref{fig:data_model_mass_comparisons} shows that 
the masses predicted by \XGBoosts tend to be lower, on average,
than those determined by \HAMs for all but the lowest \Mhams masses. 
This is in agreement with Figure~\ref{fig:hod_dv_frac_diff_mass}, which
showed that the masses determined by \HAMs tend to have larger \fracd's 
than the \mpred's by \XGBoosts for $\mpredm>10^{12}\msunh$. In the case 
of \DYN, the \XGBoosts predicted masses are larger, on average, than 
those determined by \DYNs at small dynamical masses, but smaller for 
\Mdyns larger than $\Mdynm > 10^{12}\msunh$. This is also in agreement 
with what we expect based on Figure~\ref{fig:hod_dv_frac_diff_mass}.
The qualitative agreement between these results from SDSS and what
we found in our mock catalogue is encouraging.

Our tests with mock catalogues suggest that these predicted halo
masses for SDSS galaxies may be significantly more accurate than those
estimated using \HAMs or \DYNs methods, especially at large masses.
Naturally, the worry with using these masses is the possibility that the
real universe does not look like our training mock data in some critical 
way and that the predicted SDSS masses thus contain a large systematic 
error. Though this is certainly possible, it is not likely because the 
mock catalogues were constructed to have several statistical properties
that are in agreement with the SDSS data. Moreover, \HAMs and \DYNs masses 
are known to have large systematic errors. We thus feel fairly confident
that our ML halo masses are the best available measurements for galaxy 
halo environments in the SDSS and are safe to use.

%% file: 07_summary_discussion.tex

\section{Summary and Discussion}
\label{sec:summary_discussion}

In this paper, we estimate halo masses of galaxies by employing
machine learning (ML) techniques, and we compare these to results 
by other, more traditional, mass estimation techniques, such 
as \textit{Halo Abundance Matching} (\HAM) and \textit{Dynamical 
Mass Estimates} (\DYN). We are motivated to explore ML because of
limitations in these traditional methods and because we expect that
we can obtain more precise halo mass estimates if we use information
from all the galaxy properties that correlate with mass, such as 
luminosities, colours, group dynamics, and large-scale environments.

We investigate three ML algorithms: \XGBoost, Random Forest (\RF), 
and neural networks (\NN). Each of the algorithms is trained on 
synthetic mock galaxy catalogues to predict the masses of galaxies' 
host halos, using a set of \textit{features} selected from 
both galaxy- and group-related properties. The mock catalogues
were constructed to have the same clustering and same distribution 
of \textit{observed} properties as the SDSS data, such as luminosity, 
\gr colour, and \ssfr. The final set of nine features that we use 
(\refsec{subsec:feat_selection}) are chosen based on their 
\textit{feature importance} towards the overall prediction of 
halo mass, i.e., how much each feature contributes to the overall 
prediction of halo mass. To quantify the performance of the ML 
algorithms, we test them using an independent set of mock catalogues
and we compare them to the \HAMs and \DYNs methods. We probe to what 
extent the trained ML models can be universally applied by testing 
them on data that have different properties from the training data. 
Specifically, we investigate variations in the halo occupation 
distribution (HOD), velocity bias for satellite galaxies, and the 
mass-luminosity relation for central galaxies. Finally, we apply our 
mock-trained \XGBoosts model to the \MD{19} galaxy sample and produce a 
SDSS catalogue that contains predicted halo masses, as well as the nine 
features used and the \HAMs and \DYNs masses.

The main results of our work are as follows:

\begin{enumerate}[label=(\roman*), leftmargin=0.5\parindent, labelsep=0.5\parindent]
    \item
    We determine the set of nine features (out of the 19 features from
    \refsec{subsec:feat_selection}) that contribute the most to the
    prediction of a galaxy's host halo mass. Among the set of nine
    features, we find that the two strongest features are the \rbands absolute 
    magnitude of the galaxy and the absolute magnitude of the brightest 
    galaxy in the group to which the galaxy belongs. Following these
    are the \gr colour and specific star formation rate of the galaxy
    and the group as a whole, the size and velocity dispersion of the
    group, and the galaxy's distance to the nearest cluster.
    \item
    We find that \HAMs and \DYNs overestimate halo masses on average for
    large \mpred, reaching average fractional errors in $\mathrm{log}M$ as 
    high as 10\% at the highest masses. This is due to group-finding errors
    that misclassify some galaxies as satellites and thus assign them too large 
    halo masses. At low \mpreds \HAMs works well, but \DYNs underestimates
    galaxies' halo masses. In contrast, the ML algorithms all predict halo 
    masses that are unbiased, on average, across the whole range of masses
    probed. To quantify the typical error in predicted halo mass, we calculate
    the quantity \deltalogm, where 68\% of galaxies have their masses predicted 
    with an error less than this. The three trained ML models have values
    for this typical mass error of $0.23-0.25$ dex and $0.51-0.60$ dex for 
    values of \mpreds smaller or greater than $10^{12.5}\msunh$, respectively.
    On the other hand, \HAMs yields typical halo mass errors of $0.27$ dex and 
    $0.90$ dex for the low-mass and high-mass regimes, respectively, while 
    \DYNs can only recover halo masses to $0.92$ dex and $1.25$ dex for low
    and high masses.
    \item
    When tested against mock data built with different assumptions than the
    training data, ML models mostly perform well. Results are insensitive to
    the presence of satellite galaxy velocity bias or the amount of scatter 
    in the mass-luminosity relation for central galaxies. When we vary the
    relation between halo mass and occupation number, there is no effect at
    large masses, but predicted masses can be over-estimated in the low mass
    regime. However, ML predictions still outperform \HAMs and \DYNs
    \item
    Predicted \XGBoosts halo masses for galaxies in the \MD{19} sample 
    are similar to \HAMs masses, but higher than \DYNs masses in the 
    low mass regime, but smaller, on average, than \HAMs or \DYNs
    masses in the high mass regime. This is in qualitative agreement 
    with our testing results on mock catalogues.
\end{enumerate}

These results demonstrate the power of using ML algorithms to infer 
the \textit{true} underlying mass of a galaxy's dark matter halo. 
Spectrophotometric properties of galaxies and their groups, dynamical
properties of the groups, and large scale environments, all correlate
with halo mass in different ways. It is thus not surprising that, when 
used jointly, they deliver tighter constraints on halo mass than any
one method. Our results confirm this, especially at large masses, where
methods like \HAMs and \DYNs suffer from the standard group-finding 
errors that mistakenly place some field galaxies into large groups.

The big caveat to these results is that they only hold to the extent
that the mock catalogues used to train the ML algorithms match the real
universe. We have taken care to make sure that our mock galaxies have 
distributions of observed properties and clustering that are consistent
with those in the SDSS. However, we cannot guarantee that the correlations 
between these properties and halo mass are correct in the training data.
Though our tests modifying the galaxy-halo connection are encouraging,
we have not explored the whole possible space of mock catalogues.
Readers are advised to use the SDSS predicted masses in 
\refsec{sec:applying_sdss} at their own discretion.

Perhaps the most interesting implication of this paper is the possibility
that we can use ML approaches to eliminate some of the systematic issues 
with the group-finding process, such as merging of galaxies from
different host haloes into the same group, or the splitting of
galaxies from the same halo into several different galaxy groups. For 
example, galaxies in the same group that have very discrepant ML-predicted 
halo masses may have been incorrectly grouped together. We plan to explore
this in future work.

%% file: 08_acknowledgements.tex

\section{Acknowledgements}
\label{sec:acknowledgements}

The mock catalogues used in this paper were produced by the LasDamas project 
(\url{http://lss.phy.vanderbilt.edu/lasdamas/}); we thank NSF XSEDE for 
providing the computational resources for LasDamas. Some of the 
computational facilities used in this project were provided by the 
Vanderbilt Advanced Computing Center for Research and Education (ACCRE). 
This project has been supported by the National Science Foundation (NSF) 
through a Career Award (AST-1151650). 
Parts of this research were conducted by the Australian Research Council 
Centre of Excellence for All Sky Astrophysics in 3 Dimensions (ASTRO 3D), 
through project number CE170100013.
This research has made use of 
NASA's Astrophysics Data System. This work made use of the IPython package 
\citep{Perez2007}, Scikit-learn \citep{McKinney2010}, SciPy 
\citep{jones_scipy_2001}, matplotlib, a Python library for publication 
quality graphics \citep{Hunter:2007}, Astropy, a community-developed 
core Python package for Astronomy \citep{AstropyCollaboration2013}, and 
NumPy \citep{VanDerWalt2011}. Funding for the SDSS and SDSS-II has been 
provided by the Alfred P. Sloan Foundation, the Participating Institutions, 
the National Science Foundation, the U.S. Department of Energy, 
the National Aeronautics and Space Administration, the 
Japanese Monbukagakusho, the Max Planck Society, and the Higher Education 
Funding Council for England. The SDSS Web Site is http://www.sdss.org/. 
The SDSS is managed by the Astrophysical Research Consortium for the 
Participating Institutions. The Participating Institutions are the 
American Museum of Natural History, Astrophysical Institute Potsdam, 
University of Basel, University of Cambridge, Case Western Reserve 
University, University of Chicago, Drexel University, Fermilab, the 
Institute for Advanced Study, the Japan Participation Group, 
Johns Hopkins University, the Joint Institute for Nuclear Astrophysics, 
the Kavli Institute for Particle Astrophysics and Cosmology, 
the Korean Scientist Group, the Chinese Academy of Sciences (LAMOST), 
Los Alamos National Laboratory, the Max-Planck-Institute for Astronomy (MPIA), 
the Max-Planck-Institute for Astrophysics (MPA), 
New Mexico State University, Ohio State University, 
University of Pittsburgh, University of Portsmouth, Princeton University, 
the United States Naval Observatory, and the University of Washington. 
These acknowledgements were compiled using the Astronomy Acknowledgement 
Generator.